\documentclass[journal=iecred,manuscript=article]{achemso}
\usepackage{fancyhdr}
\usepackage{booktabs}
\usepackage{siunitx}
\usepackage{graphicx}
\usepackage{subfigure} 
\usepackage{float} 
\usepackage{tabularx}
\usepackage{color}
\usepackage{geometry}
\usepackage{natbib}
\usepackage{mciteplus}
\usepackage{footnote}
\usepackage{multirow}
\usepackage{color,soul}
\usepackage{longtable}

\usepackage{lineno}
\usepackage{multirow}
\usepackage{amssymb}
\usepackage{amsmath}
\usepackage{mathtools}
\usepackage{ragged2e}
\usepackage{blindtext}
\usepackage{commath}
\SectionNumbersOn
\usepackage{easyReview}
\usepackage[version=3]{mhchem}

\usepackage[labelsep=period]{caption}
\usepackage{titlesec}
\titlelabel{\thetitle.\quad}
\author{Manohar Jammula}

\author{Somasekhara Goud Sontti}
\email{somasekhar.sonti@iitdh.ac.in}
\affiliation{Multiphase Flow and Microfluidics (MFM) Laboratory, Department of Chemical Engineering, Indian Institute of Technology Dharwad, Dharwad, 580011, Karnataka, India}

\title[An \textsf{achemso} demo]
  { Numerical analysis of controlled droplet formation surrounded by shear-thinning fluid in a co-flow microfluidic device}

\abbreviations{IR,NMR,UV}
\keywords{American Chemical Society, \LaTeX}

\begin{document}









\begin{abstract}
  
  In this work, we used a coupled level-set and volume-of-fluid (CLSVOF) computational method for studying the generation of microdroplets in a two-dimensional circular co-flow microfluidic device. The fundamental understanding of shear-thinning behavior on controlled droplet formation in a microchannel is systematically investigated. Droplet generation and dynamics are controlled by altering the operating conditions and fluid properties, such as CMC solution concentration, both fluids' flow rates, and interfacial tension. The droplet length, velocity, liquid film thickness, and formation frequency are quantitatively analyzed and presented in a non-dimensional form. The droplet length decreased with the flow rate and concentration of the CMC solution. However, the length of the droplet increased with dispersed phase flow rate and interfacial tension. From the set of simulation data, the flow regime maps and scalings relations are developed over a range of operating conditions. This work provides critical insights into the uniform and controlled droplet formation in shear-thinning fluids within a circular co-flow microchannel, offering guidance for future innovations in the design and optimization of microfluidic systems for specific applications.
  
  
  
\end{abstract}


\section{Introduction}


Droplet microfluidic devices play a substantial role in the domain of drug discovery\cite{debs2012functional,santana2019microfluidic}, cell biology\cite{chen2016controlled}, microencapsulation\cite{mou2018trojan}, disease diagnosis\cite{jang2017droplet}, nano particle synthesis\cite{li2017microfluidic} , biochemical analysis\cite{song2006chip}, in addition to that, droplet microfluidic devices act as microreactors\cite{guo2020continuous} for process intensification to promote high surface area\cite{elvira2013past}, heat\cite{yi2014pdms} and mass transfer\cite{jiang2023distribution,nieves2015openfoam}. A good control of droplet size, velocity, and formation is essential for potential applications such as chemical reactions and drug delivery. Most often, microdroplets are generated by an external force\cite{chong2016active} or by passive hydrodynamic pressure fluctuation\cite{zhu2017passive}. The walls of the microfluidic device act as boundaries to the immiscible microflow, which manipulates the fluid flow, resulting in generating droplets by a passive method using viscous shear force, with crossflow\cite{kumar2020pressure}, co-flow\cite{guillot2007stability}, and flow-focusing\cite{liang2019manipulable} devices being the most commonly used passive microfluidic devices. In addition to the standard microfluidic devices, the modified microfluidic devices were used to enhance the flow uniformity and fluid-fluid interactions.\cite{moorthy2024modified} \\


In a co-flow microfluidic device, the dispersed phase is pumped through the capillary, and the shearing effect of the coaxial continuous phase breaks the dispersed phase into droplets\cite{lian2019investigation}. The co-flow device can produce uniform-sized, spherical droplets among the microfluidic devices. Recently \citet{zhang2021microfluidic} used novel micro-nano 3D printing technology to fabricate the 3D co-flow microfluidic device for the generation of  oil-in-water (O/W) and water-in-oil (W/O)  droplets, the coefficient of variation of mono-sized droplets was found to less than 3\%. Their findings indicate that the developed micro 3D printing device provides a new method for stable droplet generation in oil-water systems.~ Similar to that~\citet{pan2021flow} experimentally investigated polystyrene-fluorobenzene (PS-FB) droplet formation in a 2\% Polyvinyl alcohol(PVA) using a co-flow microfluidic device, 4\% to 28\% PS-FB range of concentrations were used to examine the effect of the concentration on droplet size. Their results showed that the satellite droplets were observed at lower concentrations (4\% to 12\%) of PS-FB with increasing in size at constant outer phase flow rate. Eventually, the concentration of PS-FB was increased beyond 12\%, and then a liquid thread formation was noticed, which broke at one point and retracted without any satellite droplets. \\

\citet{zhang2020experimental} experimentally investigated the droplet formation, size, and corresponding flow regime with a normal ($0^{\circ}$) and tapered ($2.8^{\circ}$) co-flow microfluidic device. Dripping flow results showed that droplet size and frequency are equal in normal and tapered channels in the dripping regime. However, the droplet size is reduced, and frequency is increased in the tapered microchannel in the jetting regime. This explains the strong effect of the convergence angle on the droplet size and frequency. Fluid properties and operating conditions also play a significant role in achieving application-specific droplets in size and shape in passive microfluidic devices\cite{shang2017emerging}.~\citet{lian2019investigation} experimentally investigated the effects of flow rate and interfacial tension on the droplet size and formation frequency using a co-flow microfluidic device. The continuous and dispersed phases flow rates were regulated to produce the droplet size and frequency in the range of 101-1550 $\mu m$ and 0.32-100 Hz, respectively. The key finding from their work is that as the continuous phase flow rate increases due to the drag force, the droplet diameter decreases and frequency increases, and vice versa for the dispersed phase flow rate.\\


In reality, many fluids we encounter in our daily lives, including food\cite{ren2016synthesis}, chemicals\cite{stickel2005fluid}, biological\cite{hemminger2010flow}, and materials synthesis\cite{zhao2014spherical} behave in a non-Newtonian manner and exert non-homogeneous viscous and elastic force on the fluid interface, which significantly affect droplet formation and dynamics. \citet{chen2020modeling} studied the numerical investigation on the influence of rheological parameters, power law index n, and consistency coefficient K of power-law fluid on Newtonian droplet formation in a flow-focusing device. For shear- thinning fluid, $n$ varied from 0.35 to 1.0 at $K$= 0.35 $Pa.s^{n}$ and for shear thickening fluid, the $n$ varied from 1.0 to 1.5 at $K $= 0.001 $Pa.s^{n}$, and consistency coefficient $K$ varied from 0.2 to 1 at $n$ = 0.47, for both cases, the droplet length decreased with the increasing of the power law index and consistency coefficient, and as the $n$ increasing, the droplet regime shifts from a stable squeezing regime to an unstable jetting regime.  Nevertheless, the results were good for understanding the role of rheological properties on droplets. There is still a lack of fundamental understanding of non-Newtonian fluids such as controlled droplet formation. \\


An experimental investigation on droplet formation in a flow-focusing device was conducted by \citet{du2018breakup} with CMC solution as a dispersed phase and mineral oil as a continuous phase. In the growth and squeezing stages, the viscous stress of the dispersed phase resists the neck thinning formation; however, upon increasing the shear force of continuous force, the neck thinning is advanced in the stretching stage, eventually leading to pinch-off. Recent work of \citet{wang2023generation} showed the shear-thinning non-Newtonian droplet formation in a parallelized T-junction microchannel. The flow patterns, such as slug flow, transition flow, and parallel flow, were identified as affecting dispersed phase concentration and flow rate. The higher the concentration of CMC solution, the higher the viscosity, leading to a decrease in droplet size. \citet{xue2019non} experimentally studied the droplet dynamics with different non-Newtonian fluids as a dispersed phase. Droplet size is affected significantly by elasticity fluid compared to shear-thinning fluid. The satellite droplets were observed due to the shear-thinning behavior.

Even though there are productive advancements in the experimentation of microfluidics technology, there is uncertainty about the accuracy of the results at a microscale level, experiments are expensive, and some physical 
parameters cannot be quantified by experimentation. Hence, a reliable and well-validated numerical model could significantly simplify the difficulties. It is necessary to develop a powerful computational fluid dynamics (CFD) model development in order to understand the underlying physics of droplet formation in non-Newtonian systems in order to design the microfluidics and predict several parameters, including droplet size and shape, velocity, frequency of formation, flow regimes, liquid film thickness, pressure drop, and viscosity. Interface capturing is an important criterion in order to model the immiscible multiphase flow and widely used interface capturing techniques are the Lattice Boltzmann method (LBM)\cite{xu2021recent}, the phase field method\cite{bai2017three}, volume-of-fluid (VOF)\cite{saha2024numerical} and Level-set (LS)\cite{bashir2011simulations}.\\

\citet{sontti2019cfd} computationally investigated the bubble formation in a non-Newtonian fluid by modeling a two-dimensional (2D) axisymmetric co-flow microfluidic device using a VOF technique, and the bubble size, velocity, and formation frequency were determined. The scaling correlation for non-dimensional bubble length was proposed in terms of modified Capillary number. \citet{cao2024shear} also numerically studied the shear-thinning (Sodium Alginate) droplet formation in a 2D axisymmetric co-flow microfluidic device. The VOF method was implemented for interface capturing, and droplet dynamics, viscosity effects, and flow regime maps were reported from the simulation results. \citet{fatehifar2021non} numerically explored the droplet generation in a cross-junction microfluidic device using the VOF interface technique. The influence of the shear-thinning dispersed phase fluid concentration on droplet size was numerically predicted.\\


The key feature of the VOF method is the ability to model interface capture with maximum accuracy in mass conservation. However, due to the unstable pressure differences at the interface, the spurious currents developed. The LS function is highly efficient in capturing the sharp and well-defined interface, while VOF is capable of ensuring mass conservation\cite{balcazar2016coupled}. The combination of the VOF and LS (CLSVOF) approach, overcomes the deficiencies of both VOF and LS methods. It successfully tracks the interface with reduced error in mass conservation and efficiently captures a smooth interface\cite{sussman2000coupled,sontti2019numerical}. CLSVOF is effectively used in several applications, such as mixing characteristics\cite{qi2023numerical}, multi-bubble rising\cite{yu2023numerical}, bubble dynamics in a stirred tank\cite{lai2024spatial}, hydrogen bubble-liquid metal  interactions.\cite{shao2024clsvof}

~\citet{carneiro2019high} numerically investigated the droplet formation in a flow-focusing microfluidic device with VOF and CLSVOF interface tracking and compared the simulation results with experimental results. The interface with the CLSVOF 
 method had no spurious currents, and the droplet shape was almost similar to the experimentally obtained shape. \citet{mahmoudi2024hydrodynamic} numerically investigated droplet formation in a T-junction microchannel using CLSVOF method, the effect of contact angle, interfacial tension, and operating conditions on flow patterns such as dripping, slug, jetting, and annular flows are investigated, further non-dimensional scaling relations are developed. Recently, \citet{sontti2023regulation} implemented the CLSVOF model for interface capturing to regulate the droplet size with geometrical modification in a flow-focusing device. It is evident that droplet-based flow physics can be captured with CLSVOF. \\


In light of the above discussion, it is evident that the non-Newtonian fluid significantly impacts droplet size and formation. The droplet formation and dynamics in non-Newtonian fluids using microfluidic devices are still unexplored in detail. Questions such as how the rheological properties control the droplet size as well as how fluid properties affect the flow behavior (droplet velocity, and flow regime) are crucial. This work aims to provide a comprehensive analysis of controlled droplet formation in non-newtonian fluids. In this work, a systematic study is presented on how to control the droplet formation conditions and fluid properties in order to maintain controlled droplet formation. We propose a CFD model for circular co-flow microfluidic devices based on a CLSVOF interface capturing technique for analyzing droplet formation. Furthermore, this work illustrates a flow-regime maps and scaling relations for a droplet-based co-flow system designed specifically for shear-thinning fluids, which could potentially be crucial in designing a chemical reaction and drug delivery experiment.


\section{Computational Model}
\begin{figure}[!ht] 
	\centering
	\includegraphics[width=\columnwidth]{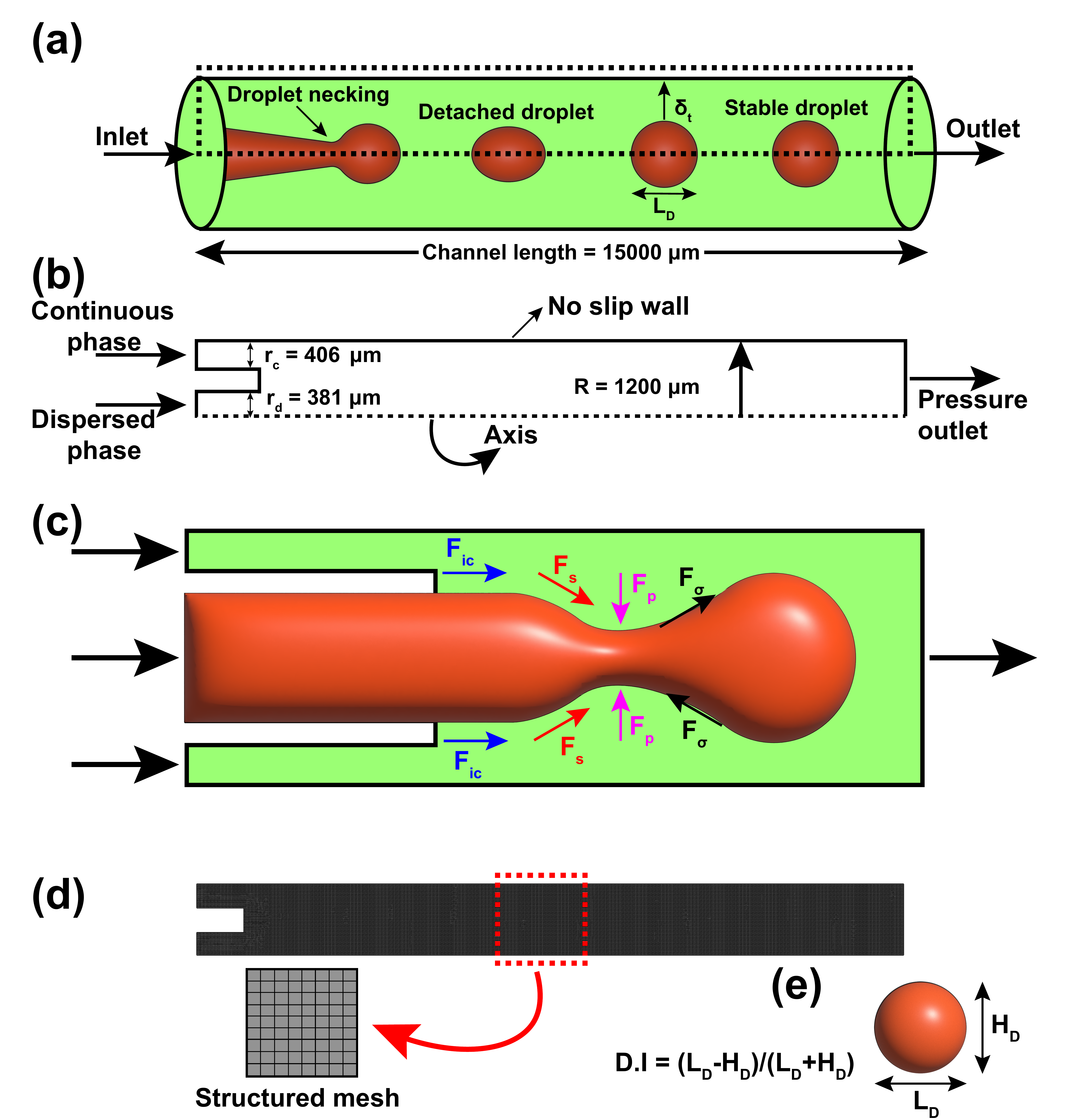} 
	\caption{ (a) Schematic of droplet formation in a circular co-flow microfluidic device where the droplet length from the rear to nose is denoted with $L_{D}$ and distance between the wall to droplet is deonted as liquid film thickness  $\delta_{t}$. (b) 2D axisymmetric  computational domain with the channel dimensions and boundary conditions. (c) The  forces acting on the dispersed phase during droplet generation in a co-flow microchannel,  (d) structured mesh of the domain, and (e) deformation index definition based on the droplet height and length. }
	\label{Figure 1}
\end{figure}

 A circular co-flow microchannel device is considered to generate the droplets in a non-Newtonian shear-thinning fluid and the dynamics of the droplets for various fluid and operating conditions. The schematic of a co-flow microchannel and the droplet formation mechanism is shown in Figure \ref{Figure 1}a.~In a co-flow microchannel, the dispersed phase enters the main channel through a small capillary inlet surrounded by a continuous phase. Both fluids meet at the end of the dispersed phase outlet in the main microchannel, and droplets are generated due to the shear force acting on the dispersed phase by the continuous phase. The forces that influence droplet formation are shown in Figure \ref{Figure 1}c, $F_{IC}$ and $F_{s}$ are inertia and shear force exerted by continuous phase on dispersed phase. As the inertia force, shear force, and pressure force ($F_{p}$) act on the dispersed phase, the interfacial tension force ($F_{\sigma}$) acts normal to interface and resist the external force on the dispersed phase. As the external force caused by the continuous phase exceed the interfacial tension force, it leads to droplet neck formation; eventually, the droplet will pinch off from the bulk of the dispersed phase. Once the droplet is generated, due to the constant flow of the dispersed and continuous phases, the droplet's movement is in the downstream direction in the microchannel.\\
 
 A 2D axis-symmetric microchannel is considered for the numerical simulation with a length (L) 15,000~$\mu$m, radius (R)  1200~$\mu$m, and radius of the continuous phase and dispersed phase are 406~$\mu$m and 381~$\mu$m, respectively as shown in the Figure \ref{Figure 1}b. The sodium carboxymethyl cellulose (CMC) solution is considered as a continuous phase, which is typically exhibits a non-Newtonian shear-thinning fluid that obeys the power law model ($\mu=K \dot{\gamma}^{\mathrm{n}-1}$)\cite{chen2020modeling}, where $K$ is consistency index, $n$ is a power law index, $\dot{\gamma}$ is a shear rate, and dispersed phase is considered as mineral oil which is a Newtonian fluid as shown in Table \ref{tab1}.\\
 \begin{table}
    \centering
    \caption{Physical properties of fluids used in the CFD simulations\cite{ma2021dynamics}.}
    \label{tab1}
 \renewcommand{\arraystretch}{1.1}
$$
\begin{array}{|c|c|c|c|c|}
\hline \text { Liquid } & \begin{array}{l}
\begin{array}{l}
\text { Density, } \rho \\
\left(\mathrm{kg} / \mathrm{m}^3\right)
\end{array}
\end{array} & \begin{array}{l}
\text { Interfacial tension, } \zeta \\
(\mathrm{mN} / \mathrm{m} \text { ) }
\end{array} & \begin{array}{l}
\begin{array}{l}
\mu_{\mathrm{c}} \text { or } K, \\
\left(\mathrm{mPa} \cdot \mathrm{s}^{\mathrm{n}}\right)
\end{array}
\end{array} & n \\
\hline \begin{array}{l}
\text { Mineral oil }
\end{array} & 876.4 & - & 22.25 & 1 \\
\hline \begin{array}{l}
0.1 \% \text { CMC in } \\
\text { water }
\end{array} & 998.9 & 3.0 & 22.682 & 0.871 \\
\hline \begin{array}{l}
0.25 \% \text { CMC in } \\
\text { water }
\end{array} & 999.7 & 2.9 & 59.079 & 0.816 \\
\hline \begin{array}{l}
0.5 \% \text { CMC in } \\
\text { water }
\end{array} & 1000.8 & 2.9 & 128.148 & 0.774 \\
\hline \begin{array}{l}
1.0 \% \text { CMC in } \\
\text { water }
\end{array} & 1002.1 & 2.9 & 220.205 & 0.717 \\
\hline
\end{array}
$$
\end{table}

 \subsection{Numerical Methodology}

In this study, both the fluids are immiscible and incompressible in the isothermal process. The fluid flow inside the channel is laminar with modified Reynold's number (ratio of inertial force to the viscous force) ranging 0.0071 to 0.8327. The droplet formation and dynamics are predicted by the solving single set of continuity and momentum equations, which are as follows: \\
Continuity equation:
\begin{equation}
\frac{\partial \rho}{\partial t}+\nabla(\rho \vec{v})=0
\end{equation}
Momentum equation:
\begin{equation}
\frac{\partial}{\partial t}(\rho \vec{v})+\nabla \cdot(\rho \vec{v} \vec{v})=-\nabla p+\nabla \cdot\left[\mu\left(\nabla \vec{v}+\nabla \vec{v}^T\right)\right]+\vec{F}
\end{equation}

Where $\vec{v}$, $\rho$, $\mu$, p and $\vec{F}$ are velocity, density, non-homogeneous viscosity of fluid, pressure, and interfacial tension force, respectively. The Bond number ($B o=\Delta \rho g r^2 / \sigma$) is less than one, so the effect of gravity is neglected in the momentum equation.

\clearpage
\subsection{Interface Capturing Methodology}
The essential feature of the CLSVOF technique is to calculate the interface's normal curvature using a smooth LS function.
The Ls function ($\Psi$) is a signed distance function, which is a function of space ($\vec{\zeta}$) and time (t). The fluid phase is recognized based on the signed changes of the LS function($\Psi$) across the interface. The LS advection is as follows  

\begin{equation}
\frac{\partial \Psi}{\partial t}+\vec{v} \cdot \nabla \Psi=0
\end{equation}

Where $\vec{v}$ and $\Psi$ are the velocity and LS function, respectively. The LS function measures the positive and negative magnitude of y, where y = y( $\vec{\zeta}$)  is the possible shortest distance from the interface to the space ($\vec{\zeta}$) in the bulk fluid at time $t$.\\

\begin{equation}
\Psi(\vec{\zeta}, t)= \begin{cases}y & \text { if } \zeta \text { is fully in the dispersed phase } \\ 0 & \text { if } \zeta \text { is in the interface  } \\ -y & \text { if } \zeta \text { is fully in the continuous phase }\end{cases}
\end{equation}

For a continuous variation of the interface, a piecewise linear interface-capturing (PLIC) scheme is used to reconstruct the interface in each cell for the succeeding time interval. 
Heaviside function [H($\Psi$)] is used to determine the density and viscosity of the dispersed and continuous phases, which are continuously varied across the interface.  

\begin{equation}
\rho(\Psi)=H(\Psi) \rho_d+(1-H(\Psi)) \rho_c ,
\end{equation}
\begin{equation}
\mu(\Psi)=H(\Psi) \mu_d+(1-H(\Psi)) \mu_c .
\end{equation}

A smoothed Heaviside function can be represented as follows
\begin{equation}
H(\Psi)= \begin{cases}0 & \text { if } \Psi<-\omega \\ \frac{1}{2}\left[1+\frac{\Psi}{\omega}+\frac{1}{\pi} \sin \left(\frac{\pi \Psi}{\omega}\right)\right] & \text { if }|\Psi| \leq \omega \\ 1 & \text { if } \Psi>\omega\end{cases}
\end{equation}

Where $\omega$ is the thickness of the interface.

\subsubsection{Surface Tension Modeling}
Surface tension force $\vec{F}$ in the momentum equation  is determined by the continuum surface force (CSF) model \cite{brackbill1992continuum} as shown below:
\begin{equation}
\vec{F}=\sigma \kappa(\Psi) \delta(\Psi) \nabla \Psi,
\end{equation}
Where $\sigma$, $\kappa(\Psi)$, and $\delta(\Psi)$ are the interfacial tension, interface curvature, and smoothed Dirac delta function.\\
 
\begin{equation}
\kappa(\Psi)=\nabla \cdot \frac{\nabla \Psi}{|\nabla \Psi|}
\end{equation}

\begin{equation}
\delta(\Psi)=\left\{\begin{array}{ll}
0 & \text { if }|\Psi| \geq \omega \\
\frac{1}{2 \omega}\left(1+\cos \left(\frac{\pi \Psi}{\omega}\right)\right) & \text { if }|\Psi|<\omega
\end{array} \right.
\end{equation}
\subsubsection{Solution methodology}

\begin{figure}[!ht] 
	\centering
	\includegraphics[width=\columnwidth]{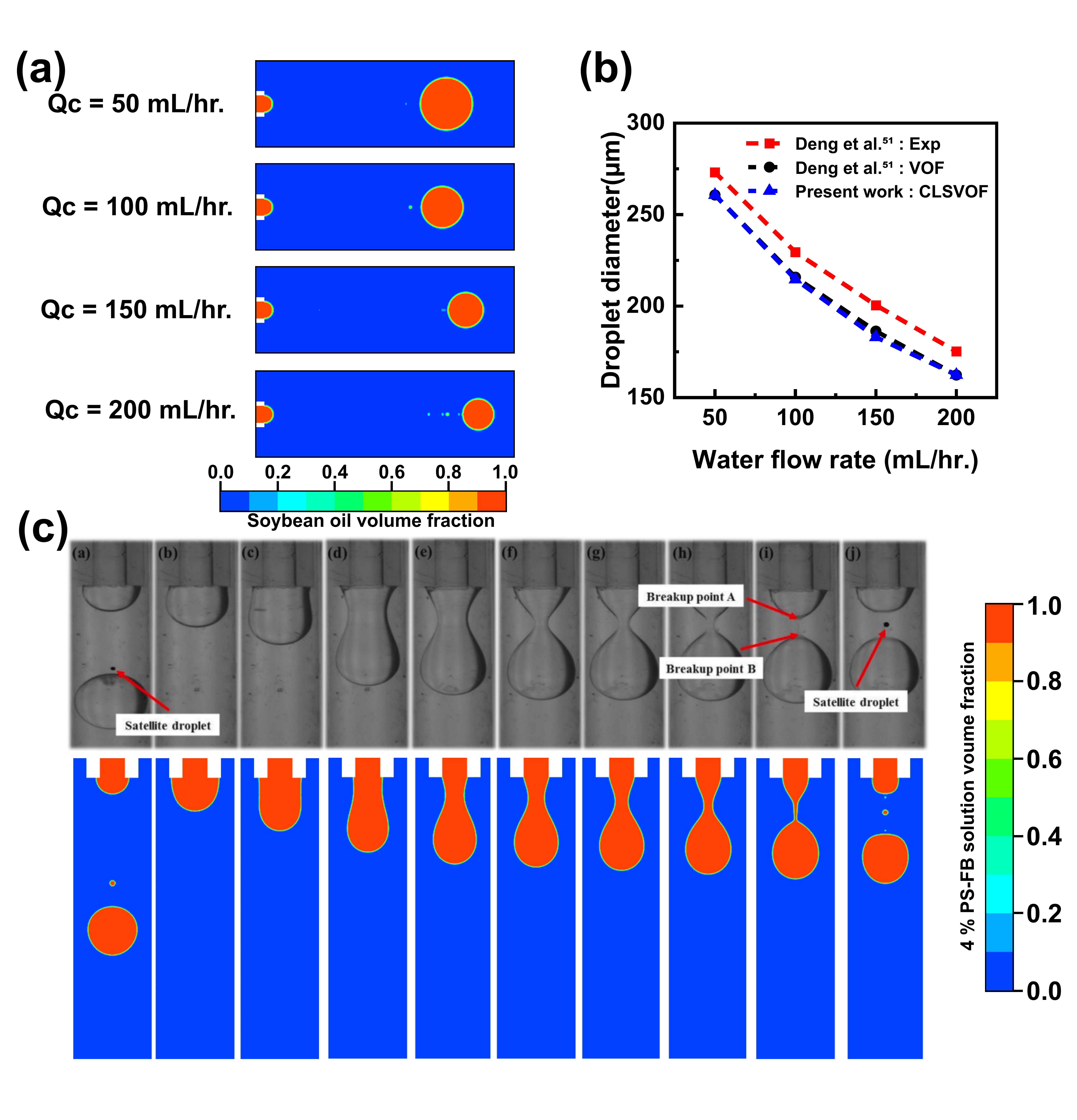} 
	\caption{Model validation (a) soybean oil volume fraction for different continuous flow rates, and (b) comparing the droplet diameter with experimental and simulation }results of \citet{deng2017numerical} at similar geometric and operating conditions. (c) Comparison of experimental droplet formation of 4\% PS-FB solution by \citet{pan2021flow} with CLSVOF model predication. Panels a-j reprinted with
permission, Copyright 2021 Elsevier~\citet{pan2021flow}.
	\label{f02}
\end{figure}

A time-dependent aforementioned partial differential equations are solved using a finite volume-based commercial Ansys Fluent CFD solver. The semi-implicit method for pressure-linked equations (SIMPLE) numerical algorithm is used to couple the pressure velocity in the momentum equation. The second-order upwind scheme is used for the spatial discretization of the momentum equation, and the first-order upwind scheme is used for the discretizing LS function equation. Since the 2D axisymmetric model is simplistic, the second-order scheme did not have a significant impact on droplet length and velocity, as compared with the first-order LS function scheme as shown in Figure S1. The piecewise linear interface construction (PLIC) algorithm is used to reconstruct the interface to predict the volume fraction in each control volume cell in the next time step. A fixed Courant number (the ratio between fluid distance to the cell distance, Co = 0.25) is considered to balance the numerical stability, accuracy, and effective capturing of the flow dynamics while limiting the numerical diffusion~\cite{hirt1981volume}. In the axisymmetric 2D simulation, the no-slip boundary condition is assumed at the wall. However, which may be differ from the experiment conditions. The constant velocity boundary condition is used at the inlets of both the continuous and dispersed phases. At the microchannel outlet, the pressure boundary condition is specified as shown in Figure\ref{Figure 1}b. It is assumed that the continuous phase completely wets the microchannel wall, and the three-phase contact angle with the microchannel wall specified is 180\textdegree. The droplet and the liquid film thickness around the droplet to the solid wall are successfully captured for the range of operating conditions reported in the present study. Therefore, the role of the three-phase contact angle on the droplet formation and dynamics is negligible. 

\subsubsection{Grid Independence and Model Validation  }
\begin{figure}[!ht] 
	\centering
	\includegraphics[width=\columnwidth]{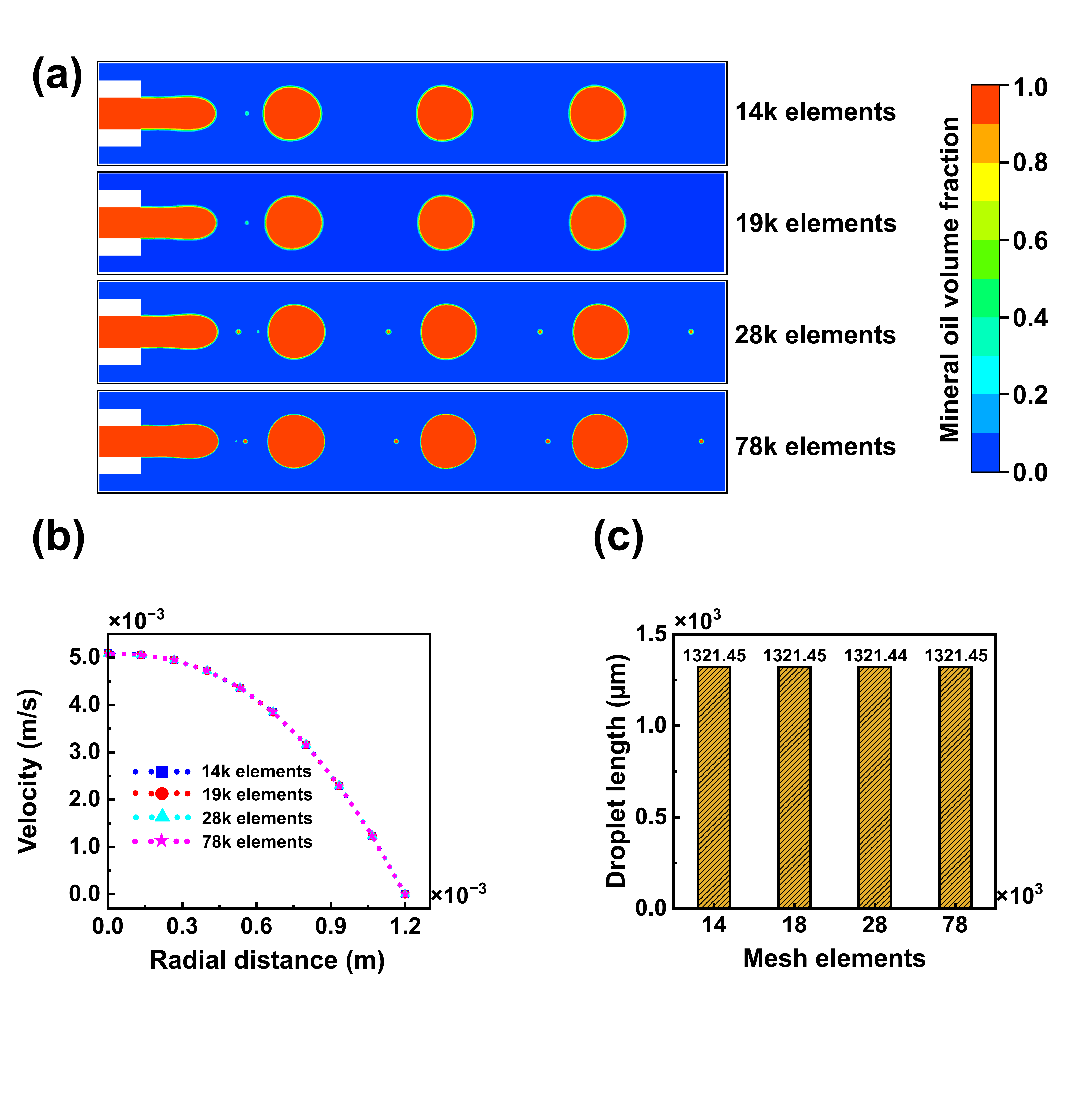} 
	\caption{Grid independence (a) volume fraction contours for different mesh elements, (b) radial velocity profile for different mesh elements, and (c) droplet size for different mesh elements of CMC-1\% solution at $Q_{c}$ = 4 mL/hr and $Q_{d}$ = 3 mL/hr.}
	\label{f03}
\end{figure}
To examine the efficacy of the developed CLSVOF model, we have quantitatively validated the soyabean oil droplet diameter in a co-flow microchannel. The experimental results of \citet{deng2017numerical} are numerically performed with the same dimensions of the microchannel and fluid properties. The soybean droplet diameter ($\mu$m) changes with deionized water flow rate are quantified as shown in Figure \ref{f02}a. Further, the CFD model predication are compared with experimental results as shown in Figure \ref{f02}b. It is evident from Figure \ref{f02}b that the numerical results are in good agreement with experimental results with a maximum deviation of less than 10\% and our simulation results are good agreement with the \citet{deng2017numerical} numerical results with a less than  1\% error. The droplet diameter results obtained using the VOF method are nearly identical to CLSVOF method results. However, the noticeable differences in interface sharpness is observed compared with VOF results. Therefore, considering the sharp interface between the two-phase, CLSVOF method is employed. In addition, the CLSVOF model is also validated qualitatively with the experimental results of the droplet formation mechanism of Polystyrene-Flourobenzene(4\% PS-FB) in a co-flow microchannel by Pan et al.\cite{pan2021flow}. The qualitative 4\% PS-FB droplet formation and volume fraction validation is shown in Figure \ref{f02}c, and the simulation results are in excellent agreement with the experimental results and the liquid film thickness in panels a and j of experimental results was compared to simulation results, showing an error of 7\%  and 6\%, respectively. Further, the model is considered for the numerical investigation of mineral oil droplet generation and dynamics by altering the physical parameters. Four different concentrations of CMC solutions are considered, and the fluid properties are tabulated in Table~\ref{tab1}.

At first, we systematically investigated the grid analysis with four different mesh elements to analyze the accuracy of the results and interface resolution. Structural meshing is considered, as shown in Figure \ref{Figure 1}d, with mesh elements 14000, 19000, 28000, and 78000, and corresponding mesh element sizes are 35 $\mu m$, 30 $\mu m$, 25 $\mu m$, and 15 $\mu m$, respectively. Interface resolution for all four mesh elements is shown in Figure \ref{f03}a. The interface resolution is relatively poor incase of  14000 and 19000 mesh elements without satellite droplets. Eventually, the interface resolution is improved upon increasing the mesh elements, and satellite droplets are also observed for 28000 and 78000 elements. This may be due to the larger mesh elements capturing satellite droplets in the system. Although these droplets can significantly impact heat and mass transfer and mixing characteristics. However, the present study shows that satellite droplet formation has negligible effects on hydrodynamics. The formation of satellite droplets can be omitted by considering a inflation layer near the axis during mesh generation. This approach is an a effective in reducing satellite droplet formation within the flow system. In spite of this, satellite droplets may not have an impact on droplet velocity or other flow conditions due to their relatively small size. The velocity profile shown in Figure \ref{f03}b and droplet length ($\mu$m) shown in Figure \ref{f03}c are the same for all the mesh elements. Considering the interface resolution, satellite droplet formation in practical situations, and simulation time, further parametric studies proceeded with mesh elements 28000.

\section{Results and Discussion}
In this work, mineral oil droplet generation is numerically examined. In accordance with the droplet formation, its dynamics are systematically investigated by varying the concentration of the continuous phase, the flow rate of the continuous phase, the flow rate of the dispersed phase, and interfacial tension.

\subsection{Effect of Continuous phase concentration}
The concentration of the continuous phase plays a significant role in the formation of droplets. Four different concentrations of CMC solutions are considered to study the effect of continuous phase concentration, as shown in Table~\ref{tab1}. As the concentration of the CMC solution increases, the corresponding effective viscosity is analytically calculated using Equation \ref{eqn11}.
\begin{equation}
\mu_{\text {eff }}=K\left(\frac{3 n+1}{4 n}\right)^n\left(\frac{8 v}{r_c}\right)^{n-1}
\label{eqn11}
\end{equation}
\begin{figure}[!ht] 
	\centering
	\includegraphics[width=\columnwidth]{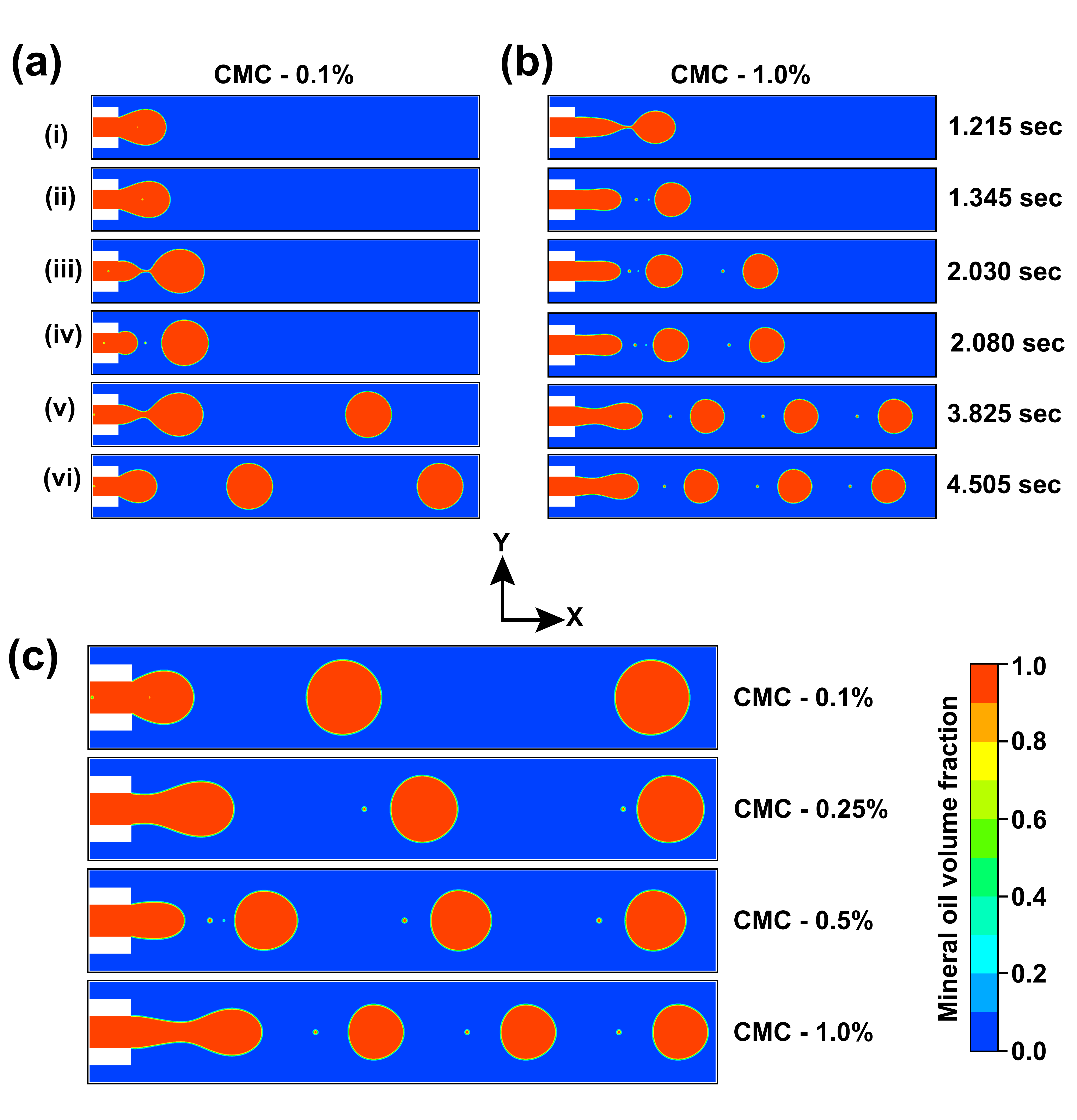} 
	\caption{Effect of concentration on (a) droplet evolution for CMC-0.1\%, (b) droplet formation mechanism for CMC-1\%, and (c) droplet volume fraction for all CMC concentrations at $Q_{c}$ = 4 mL/hr and $Q_{d}$ = 3 mL/hr.}
	\label{Figure 4}
\end{figure}
Where K is the consistency index (Pa.$s^n$), n is the power law index, $v$ is the velocity (m/s) of the fluid, and $r_{c}$ is the radius (m) of the continuous phase inlet.
In this section, the impact of the viscosity of the continuous phase is systematically investigated on the droplet formation and its dynamics at constant operating conditions. Droplet generation in mineral oil- CMC -0.1\% and mineral oil- CMC-1\% are illustrated in Figure \ref{Figure 4}. The droplet formation mechanism consists of four stages: (i) expanding, (ii) necking, (iii) pinch off, and (iv) stable droplet. 
The dispersed phase entering into the microchannel through the capillary meets the continuous phase at the exit of the capillary. For mineral oil - CMC -0.1\%, the dispersed phase starts to expand in the microchannel in the axial and radial directions as a balloon-shaped shape as shown in Figure \ref{Figure 4}a at 1.215 s. The dispersed phase continues to expand further in the microchannel as shown in Figure \ref{Figure 4}a at 1.345 s, due to the rupture caused by the shear force of the continuous phase,  the dispersed phase starts to form a thin filament which is called necking, shown in Figure \ref{Figure 4}a at 2.030 s which is eventually detached and form a droplet as shown in the Figure \ref{Figure 4}a at 2.080 s. The droplet formation for mineral oil- CMC-1\% explored under the same operating conditions, the dispersed phase expansion, and the formation of thin filaments are faster, as shown in Figure  \ref{Figure 4}b at 1.215 s. An early droplet formation is observed at 1.345 s as shown in Figure \ref{Figure 4}b. In addition, stable droplets were observed to form as the flow time progressed, as shown in Figure \ref{Figure 4}a and b at time 4.505 s. It is evident that the droplet formation rate is increasing as continuous phase viscosity increases. A corresponding mineral oil - CMC droplet illustration is shown in Figure \ref{Figure 4}c.

\begin{figure}[!ht] 
	\centering
	\includegraphics[width=\columnwidth]{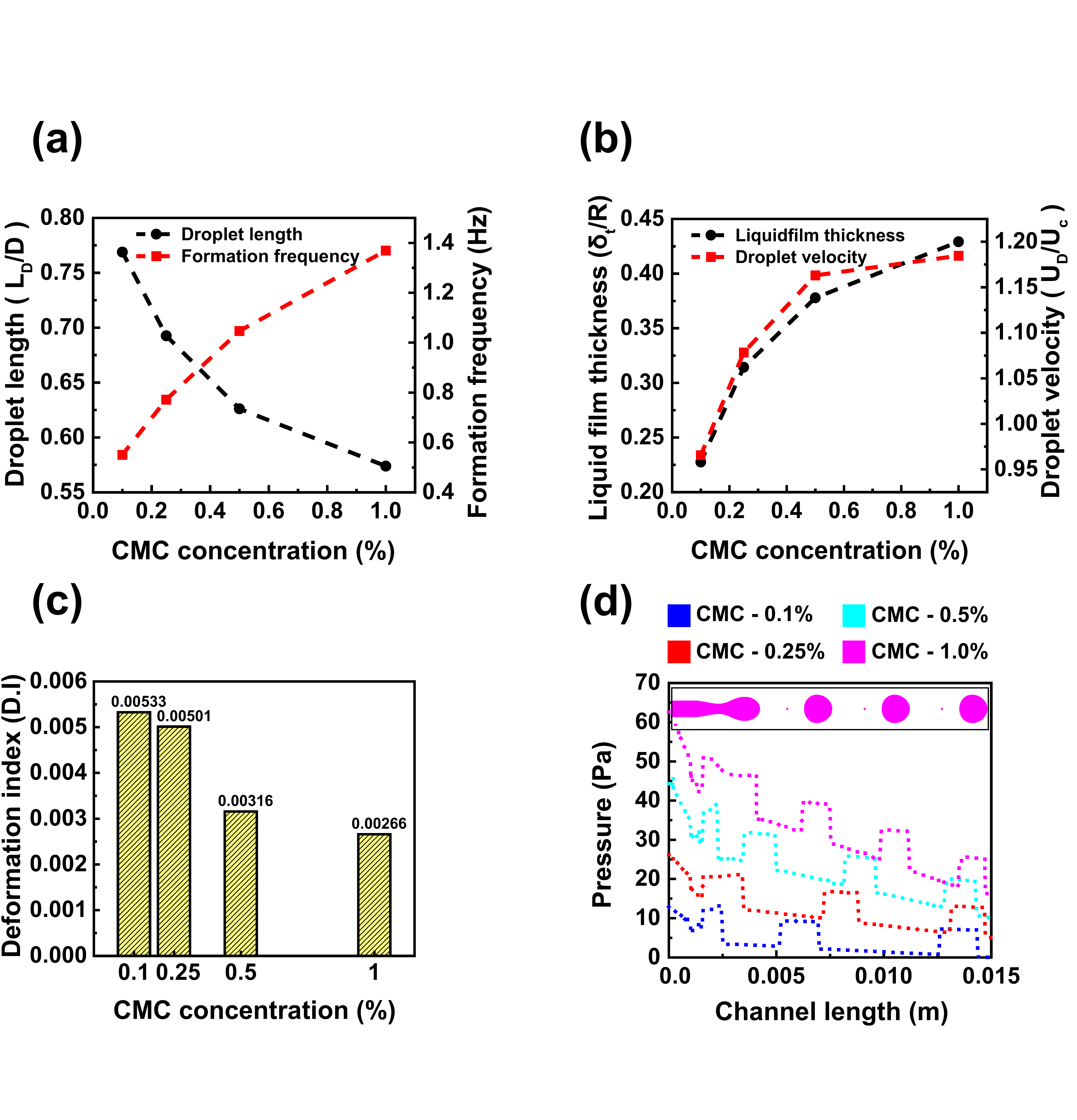} 
	\caption{ Effect of concentration on (a) non-dimensional droplet length and droplet frequency, (b) non-dimensional liquid film thickness and droplet velocity, (c)droplet deformation index, and (d) pressure profile along the microchannel length at $Q_{c}$ = 4 mL/hr and $Q_{d}$ = 3 mL/hr.}
	\label{Fig.5}
\end{figure}

\begin{figure}[!ht] 
	\centering
	\includegraphics[width=\columnwidth]{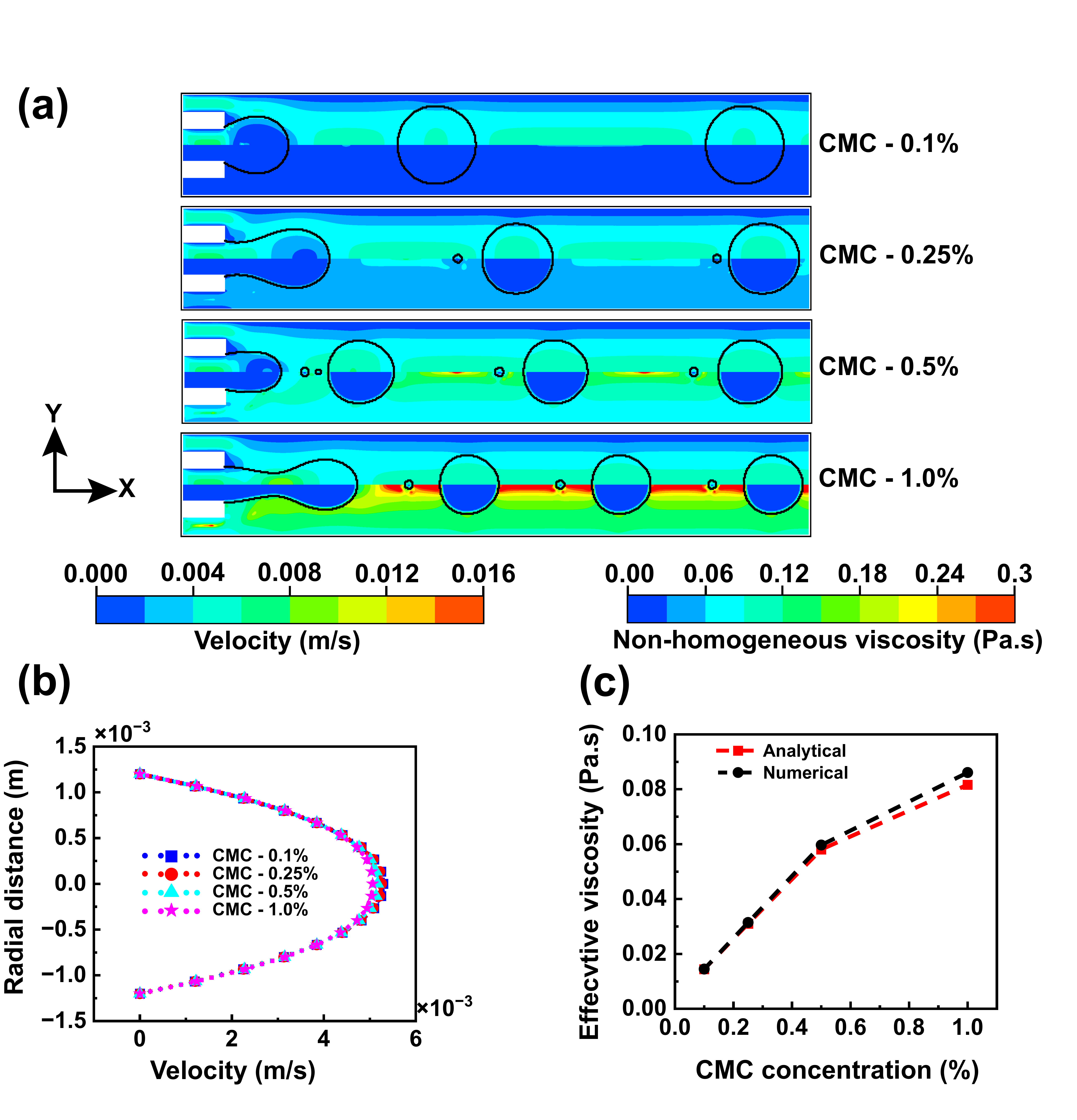} 
	\caption{ Effect of concentration on (a) velocity and non-homogeneous viscosity contours from CMC-0.1\% to CMC-1\%, (b) velocity profile in the radial direction for CMC-0.1\% to CMC-1\%, and (c) comparison of effective viscosity of CMC solution at $Q_{c}$ = 4 mL/hr and $Q_{d}$ = 3 mL/hr.}
	\label{Fig06}
\end{figure}
The mineral oil droplet length is represented as a non-dimensional droplet length (ratio of droplet length to the diameter of the microchannel, $\frac{L_{D}}{D}$) with respect to the CMC concentration in Figure \ref{Fig.5}a. It is evident that the droplet length continuously decreases as the concentration of CMC increases, and the droplet formation frequency increases with CMC concentration, as shown in Figure \ref{Fig.5}a. Liquid film thickness is the space occupied by the continuous phase between the droplet and microchannel wall. As the concentration of the CMC solution increases, shear force also increase, reducing droplet size. This reduction in droplet size creates a larger gap between the droplet and the channel wall, resulting in an increase in the liquid film thickness. A thicker liquid film allows the continuous phase to exert higher inertial force on the droplet, which reduces the resistance experienced by the droplet as it moves downstream. Consequently, the droplet is able to accelerate, resulting in a higher velocity, as shown in Figure \ref{Fig.5}b.\\

The variation in droplet size is represented by the deformation index (D.I), which is a function of the droplet's length and height, as shown in Figure \ref{Figure 1}e. As the concentration of the CMC solution increases, the deformation index decreases continuously, as illustrated in Figure~\ref{Fig.5}c. The pressure profile along the microchannel length is represented in Figure \ref{Fig.5}d. The pressure values depicted are point values measured along length of microchannel. The continuous dotted line represents the pressure profile, which shows the pressure is continuously decreasing along the channel length and also with continuous phase viscosity. In each pressure profile, there is a peak with some width, which indicates the droplet shape, its location in the microchannel, and the corresponding pressure drop between the front and rear of the droplet. The pressure directly may influences the shear forces acting on the interface between the continuous and dispersed phases. The higher pressure gradients across the channel induce stronger shear forces, which facilitate the deformation and eventual breakup of droplets. During the initial stages of droplet formation, localized pressure variations near the droplet interface can result in shape distortion, leading to changes in the length and width of the droplet.

Figure \ref{Fig06}a shows each CMC concentration's velocity and non-homogeneous viscosity contour. The top portion of the contour demonstrates the velocity, and the bottom portion demonstrates the non-homogeneous viscosity. It is corroborated by Figure \ref{Fig06}a, as the concentration of CMC solution increases, the non-homogeneous viscosity increases. The satellite droplets are clearly observed during the droplet formation. This might be due to that at lower concentrations of CMC, the interfacial tension is sufficiently strong to resist these perturbations and prevent the formation of satellite droplets. However, at higher concentrations, the viscosity of the fluid increases, and the shear force exerted on the liquid thread become more dominant than the interfacial tension. This imbalance between shear and interfacial tension forces makes the thinning liquid thread more prone to instability, resulting in the formation of satellite droplets. Figure \ref{Fig06}b demonstrates the velocity distribution in the radial direction in the microchannel, and it is evident that the velocity in the radial direction for lower concentration (CMC-0.1\%) is higher, and for higher concentration (CMC-1\%) the velocity is lower due to the high viscosity. Figure \ref{Fig06}c shows the effective viscosity of the CMC solution with respect to the CMC concentrations. The analytical effective viscosity calculated using Equation \ref{eqn11} is compared with simulation results. The simulated effective viscosity agrees with analytical viscosity with less than 5\% error.

\subsection{Effect of Continuous Phase Flow Rate}
The effect of the continuous phase flow rate of the four different CMC concentration solutions on droplet formation, length, and velocity are sequentially explored in this section. The flow rate of the dispersed phase is remained constant by varying the flow rate of all four CMC concentration solutions. In this section, we represented the flow rate in dimensionless form as a flow rate ratio ($Q^* = Q_{c}/Q_{d}$) where $Q_{c}$ is the continuous phase flow rate at a fixed dispersed phase flow rate $Q_{d}$ for quantifying the aforementioned results . 

\begin{figure}[!ht] 
	\centering
	\includegraphics[width=\columnwidth]{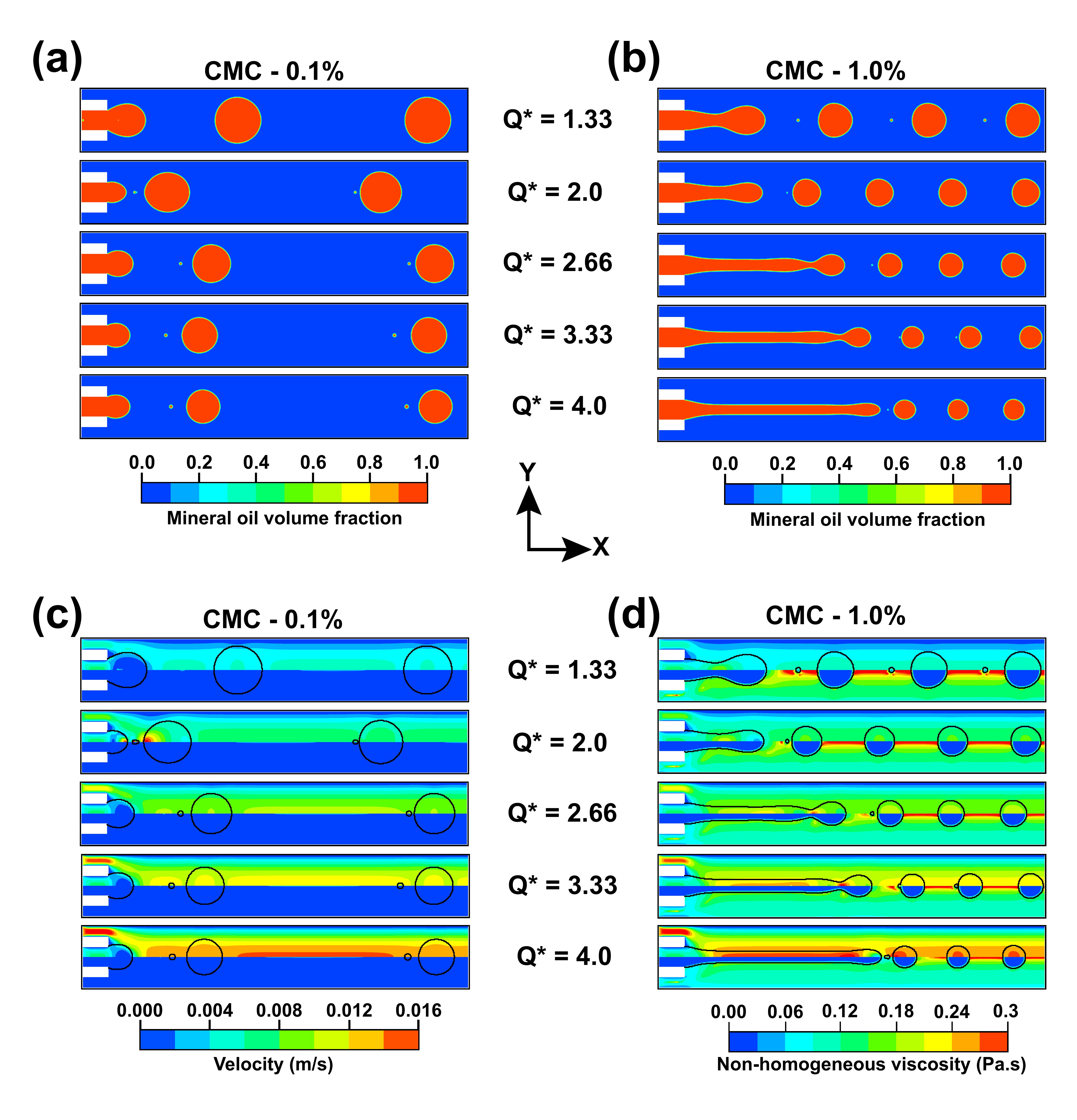} 
	\caption{ Effect of continuous phase flow rate on (a) volume fraction of mineral oil droplet in a CMC-0.1\% solution, (b) volume fraction of  mineral oil droplet in a CMC-1\% solution, (c) velocity and viscosity  contour of CMC - 0.1\%, and (d) velocity and viscosity profile of CMC - 1\% at fixed $Q_{d}$ = 3 mL/hr.}
	\label{f07}
\end{figure}

Figure \ref{f07}a shows  the mineral oil droplet formation in CMC-0.1\%,  for all the continuous phase flow rates, the droplet formation is observed at the entrance of the main channel, whereas for CMC-1\%, due to increases shear force and as the continuous phase flow rate increases the dispersed phase stretches into the channel, becomes thinner, and droplet formation position moves in the downstream direction of the main channel  as shown in Figure \ref{f07}b.  

However, for both CMC-0.1\% and CMC-1\%, the droplet size is decreases with a continuous phase flow rate due to the inertial force acting on the dispersed phase. Figure \ref{f07}c shows the velocity and non-homogeneous viscosity contour of CMC-0.1\%. A top portion of the contour shows a flow velocity field, while a bottom portion shows non-homogeneous viscosity. The color change from blue to red is noticed in the top portion of the velocity contour as the continuous phase flow rate increases ($Q^{*}$ = 1.33 to 4.0 ), whereas there is negligible change in the non-homogeneous viscosity for CMC-0.1\%. The Figure \ref{f07}d shows the velocity and non-homogeneous viscosity contour of CMC-1\% with respect to the flow rate ratio ($Q^{*}$), the velocity profile in the contour has noticeble color change as the flow rate increases from $Q^{*}$ = 1.33 to 4.0. However in the Figure \ref{f07}d it is observed that the non-homogeneous viscosity of CMC - 1\% is decreases as the flow rate is increases from $Q^{*}$ = 1.33 to 4.0 due to the inertial force are dominating than the viscous force.\\
\begin{figure}[!ht] 
	\centering
	\includegraphics[width=\columnwidth]{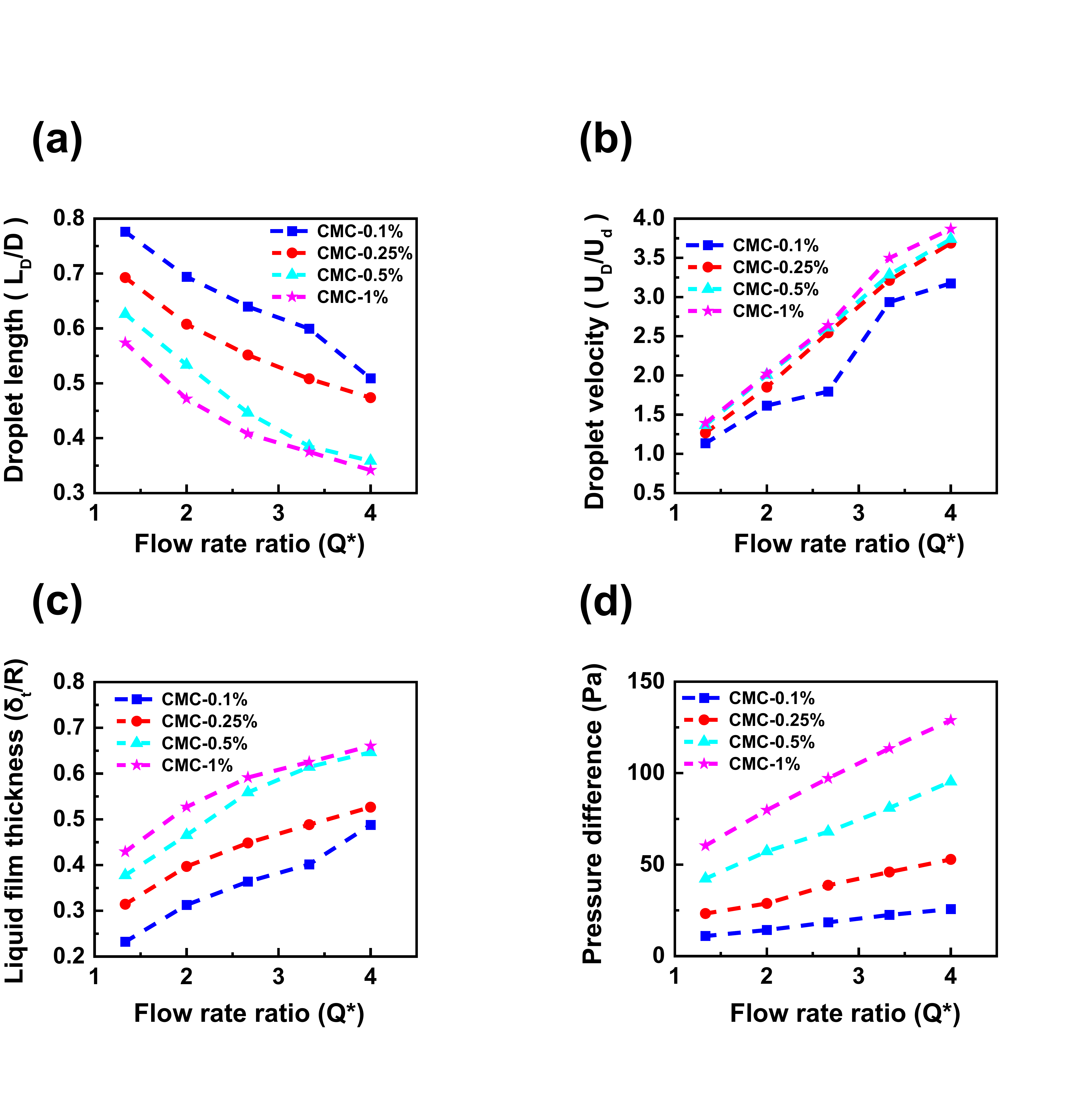} 
	\caption{ Effect of continuous phase flow rate on  (a) non-dimensional droplet length, (b) non-dimensional droplet velocity, (c) non-dimensional liquid film thickness, and (d) Pressure drop at fixed $Q_{d}$ = 3 mL/hr. }
	\label{f08}
\end{figure}
In Figure \ref{f08}, it is observed that the effect of continuous phase flow on droplet length, velocity, liquid film thickness, and pressure variation are represented with respect to flow rate ratio ($Q^{*}$). As the continuous flow rate is increases, the non-dimensional droplet length ($\frac{L_{D}}{D}$) is decreases for all the CMC concentration solutions shown in Figure \ref{f08}a, due to the inertial force and viscous force, the \% change in droplet length is higher in the CMC-0.5\% and CMC-1\%. As the droplet length is decreases with continuous phase flow rate, the space occupied by the continuous phase around the droplets is significant, which results in the increases droplet velocity represented as non-dimensional velocity $(U_{D}/U_{d}) $ with respect to flow rate ratio ($Q^{*}$) in Figure \ref{f08}b and the liquid film thickness between droplet and microchannel wall is increases which is represented as non-dimensional liquid film thickness ($\delta_{t}/R$) shown in Figure \ref{f08}c. Figure \ref{f08}d shows the pressure difference between the continuous phase inlet and microchannel outlet with respect to the flow rate ratio. As the continuous phase flow rate increases for all the CMC concentration solutions, as per the Hagen-Poiseuille equation ($\Delta P=\frac{8 \mu Q L}{ \pi r^4}$), for laminar flow in a circular channel, the pressure difference is directly proportional to the fluid's viscosity and flow rate. Therefore, as the concentration and flow rate of the CMC solution increase, the pressure difference along the channel length increases. This hydrodynamic pressure difference directly impacts droplet formation by increasing the droplet formation frequency and velocity, while reducing droplet size.  However, the pressure difference is significant in the higher CMC concentrations.

\begin{figure}[!ht] 
	\centering
	\includegraphics[width=\columnwidth]{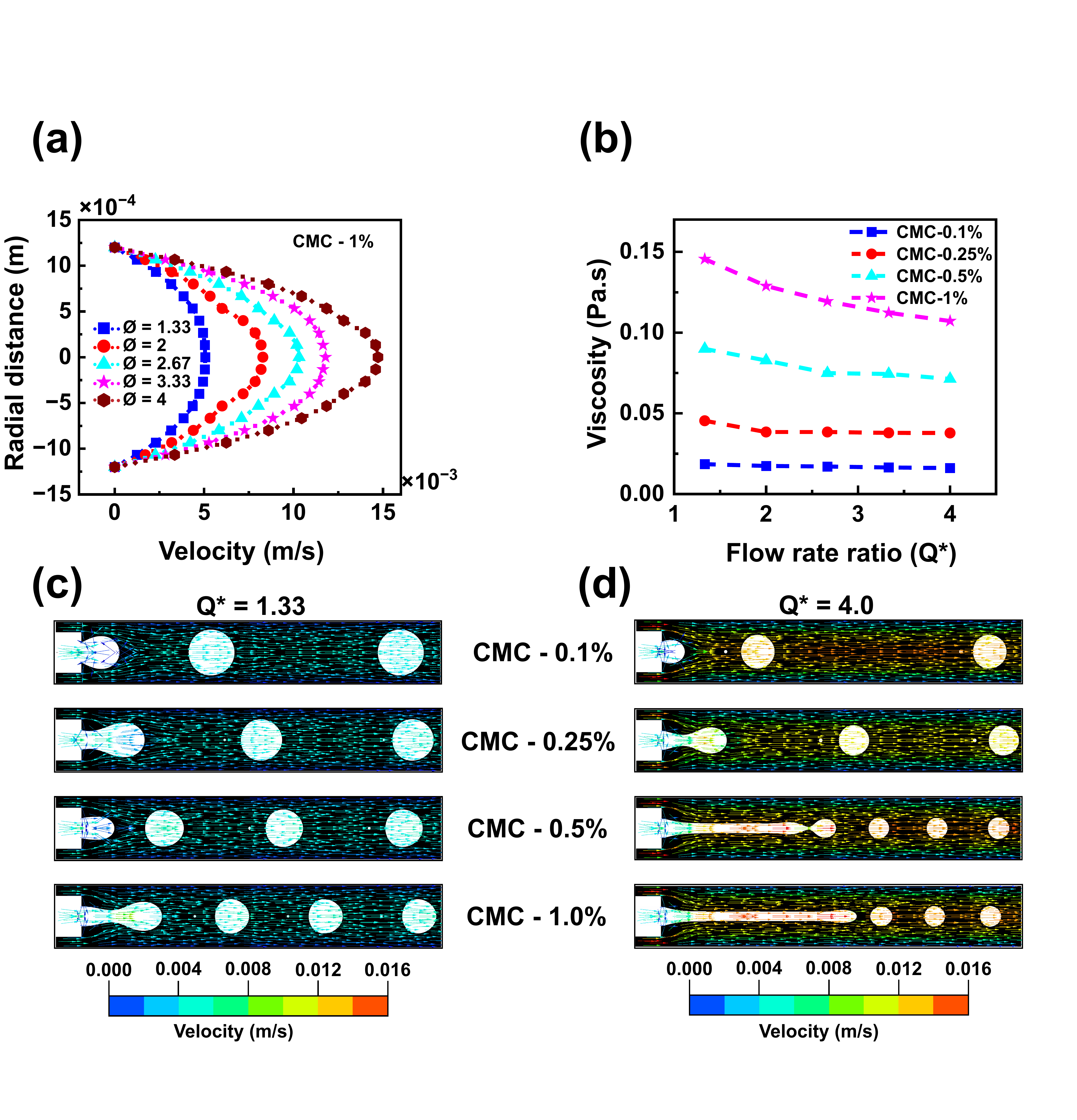} 
	\caption{Effect of continuous phase flow rate on (a) radial velocity, (b) non-homogeneous viscosity, (c) velocity vector contours at $Q_{c}$ = 4 mL/hr, and  (d) velocity vector contours at $Q_{c}$ = 12 mL/hr at fixed $Q_{d}$ = 3 mL/hr.}
	\label{f09}
\end{figure}
Figure \ref{f09}a shows the radial velocity profile with continuous phase flow rate for CMC-1\% solution. The velocity is zero at the channel wall due to the no-slip boundary condition, and velocity is highest in the middle of the microchannel. However, there is a gradual increase in the velocity from the channel microchannel wall to the center of the microchannel for all the continuous phase flow rates. The non-homogeneous viscosity of the CMC solution with respect to the continuous phase flow rate is examined. The flow rate of the CMC solution increases, leading to the imposition of a high shear rate. As the shear rate increases, the viscosity of the shear-thinning fluid decreases. However, the non-homogeneous viscosity of the CMC-1\% solution is  decreases significantly than the other CMC solutions because CMC-1\% is close to the shear-thinning non-Newtonian fluid, shown in Figure \ref{f09}b.

Figure \ref{f09}c and d shows the velocity distribution inside the microchannel. As shown, the velocity magnitude is highest along the central axis of the channel and decreases near the walls due to the no-slip boundary condition. As the flow rate ratio ($Q^*$) increases across all CMC concentrations, the velocity magnitude also increases, which is indicated by the change in the color gradient of the contours. Additionally, the flow contours illustrate significant changes in droplet size, droplet pinch-off position, and the length of the dispersed-phase thread in relation to variations in the flow rate ratio. At a lower flow rate for all the CMC concentrations, the droplet necking and pinch-off started at the channel entrance, resulting in droplet formation at the microchannel's beginning. whereas at higher flow rates, due to higher inertial force, the droplet pinch-off position for CMC -0.5\% and CMC-1\% the droplet pinch-off positioned changed to the middle of the microchannel.

\subsection{Effect of dispersed phase flow rate}
\begin{figure}[!ht] 
	\centering
	\includegraphics[width=\columnwidth]{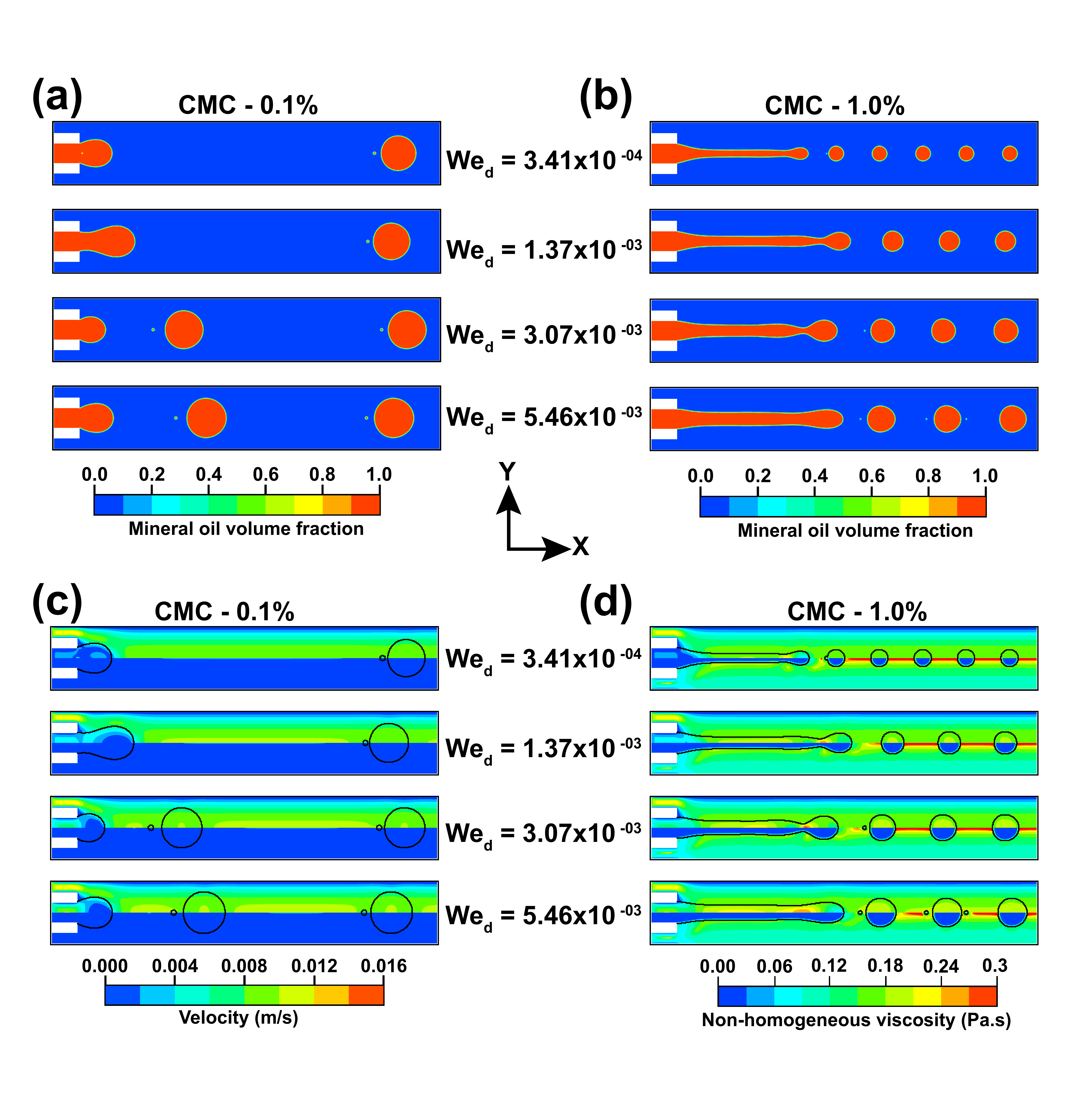} 
	\caption{Effect of dispersed phase flow rate on (a) mineral oil volume fraction in a CMC-0.1\% solution, (b) mineral oil volume fraction in a CMC-1\%, (c) velocity and viscosity contours in a CMC - 0.1\% solution, (d) velocity and viscosity contours in a CMC - 1\% solution at a fixed $Q_{c}$ = 8mL/hr.}
	\label{f10}
\end{figure}
This section systematically explores the effect of the dispersed phase flow rate on droplet formation, size, and velocity. The flow rate of the continuous phase is constant, and the dispersed phase flow rate is varied. The dispersed phase flow rate in this section is expressed in terms of non-dimensional Weber number (ratio of inertial force to the interfacial tension force, $\mathrm{We}=\frac{\rho u_{d}^2 r_{d}}{\sigma}$). The continuous phase exerts a shear force on the dispersed phase when both phases contact each other in the main channel. As the flow rate of dispersed phase increases, it expands in all directions before creating a droplet with a constant continuous phase flow rate. As the dispersed flow rate increases, the shear force required to pinch off droplets from the bulk dispersed phase is insufficient, leading to a bigger droplet size. However, the droplet generation occurred at the beginning of the microchannel for CMC - 0.1\% as shown in Figure \ref{f10}a whereas the droplet pinch-off for CMC - 1\% solution shifted to the middle of the microchannel with increasing thickness of the droplet thread shown in Figure \ref{f10}b.\\
Figure \ref{f10}c and d show the velocity and viscosity contours of CMC-0.1\% and CMC-1\%, respectively. The top portion of the contour is velocity, and the bottom is the viscosity contour. The droplet size increases with the dispersed phase flow rate, resulting in the magnitude of the velocity inside the microchannel being significantly unaffected for CMC-0.1\% in Figure \ref{f10}c. However, for CMC-1\%, due to the high shear force, it is observed that the velocity inside the microchannel is noticeably increases, shown in Figure \ref{f10}d.\\

\begin{figure}[!ht] 
	\centering
	\includegraphics[width=\columnwidth]{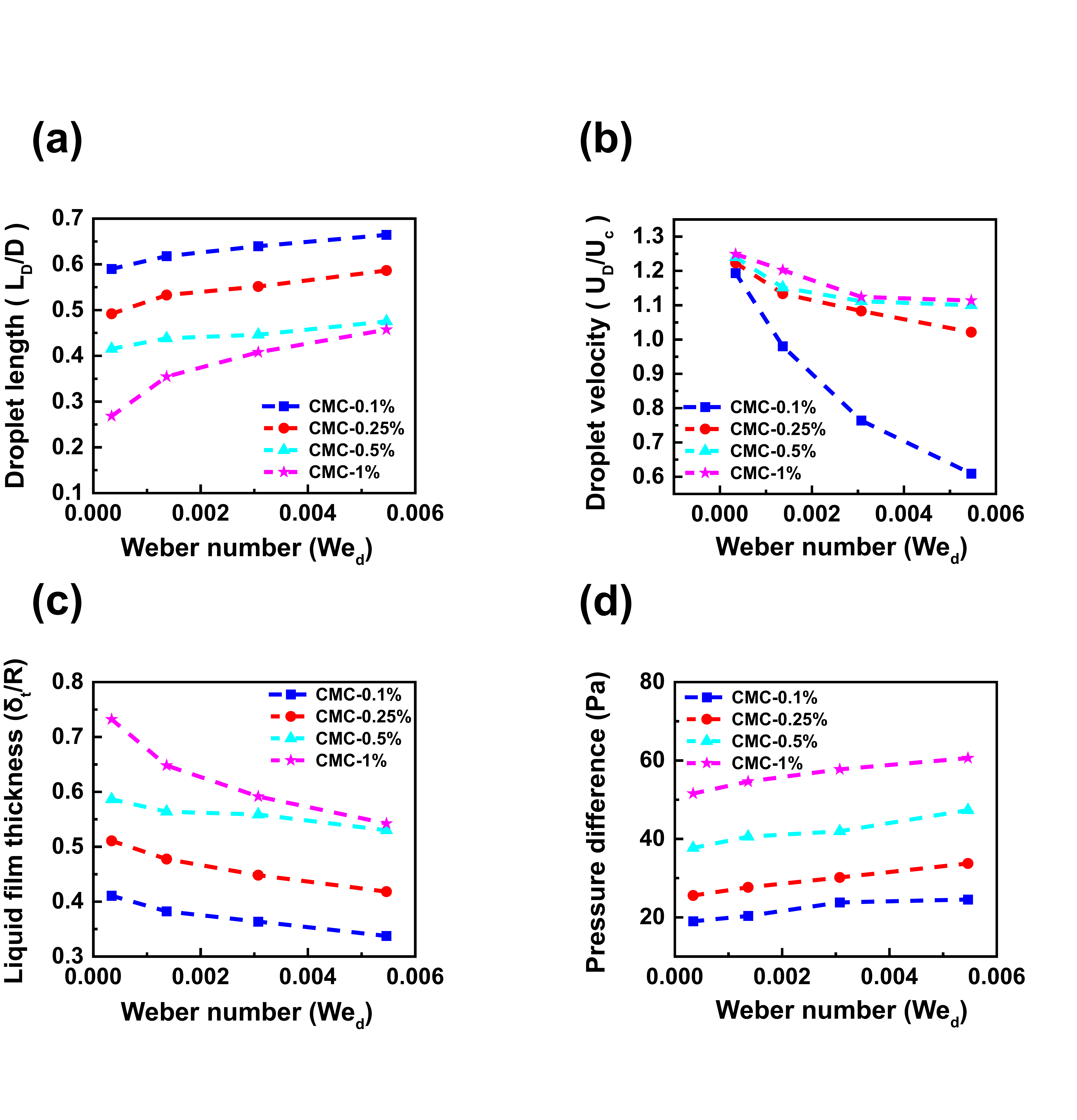} 
	\caption{ Effect of dispersed phase flow rate on (a) non-dimensional droplet length, (b) non-dimensional droplet velocity, (c) non-dimensional liquid film thickness, and (d) Pressure drop at a fixed $Q_{c}$ = 8mL/hr. }
	\label{f11}
\end{figure}

\begin{figure}[!ht] 
	\centering
	\includegraphics[width=\columnwidth]{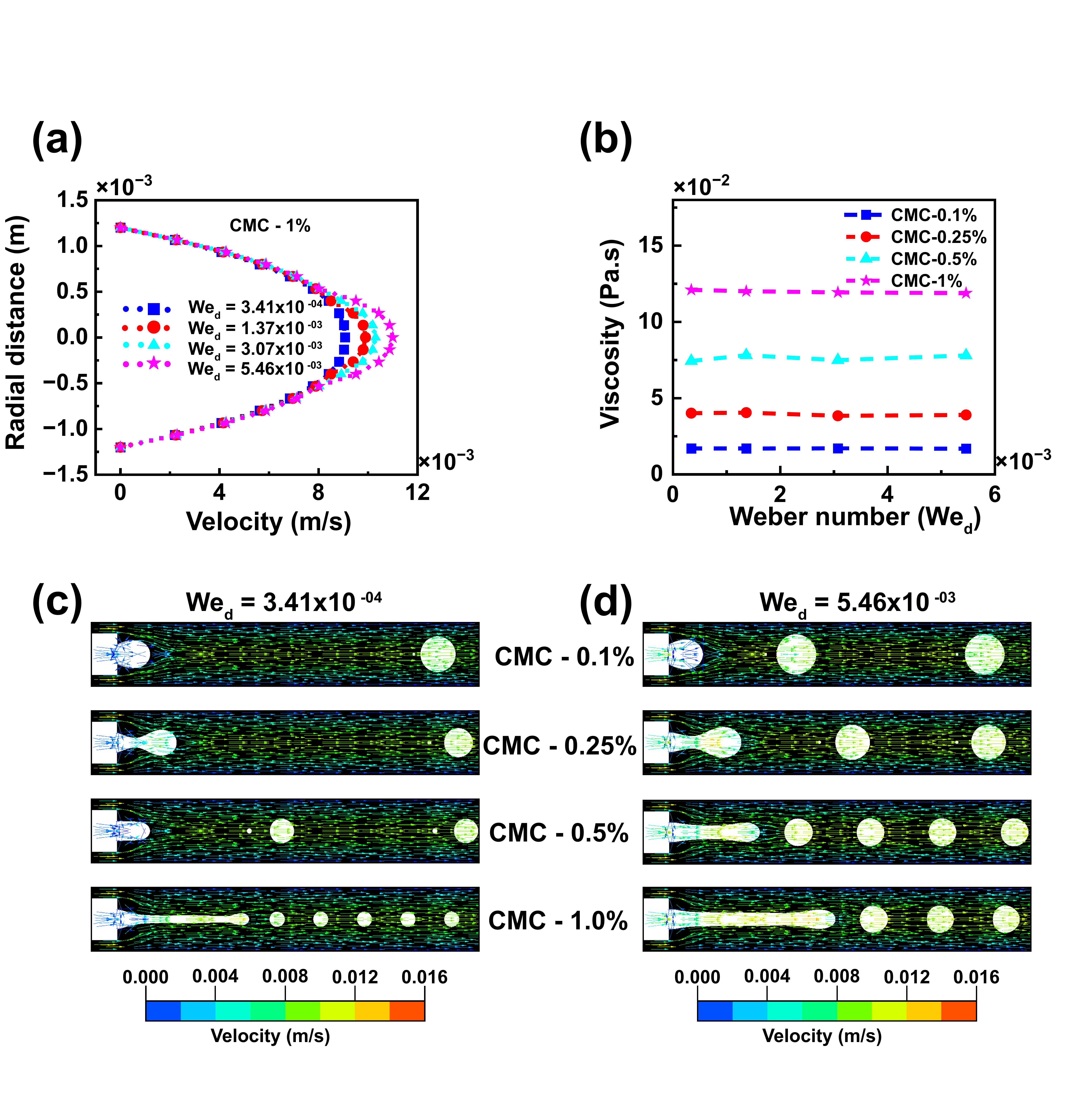} 
	\caption{Effect of dispersed phase flow rate on (a) radial velocity, (b) non-homogeneous viscosity, (c) velocity vector contours at $Q_{d}$ = 1 mL/hr, and (d) velocity vector contour at $Q_{d}$ = 4 mL/hr and  $Q_{c}$ = 8 mL/hr.}
	\label{f12}
\end{figure}
As the Weber number increases due to  increase in flow rate of the dispersed phase , the non-dimensional droplet length ($L_{D}/D$) is increases for all the CMC concentration solutions. This is due to that the larger inertial forces push against the shear force exerted by the continuous phase, causing the droplet to enlarge before pinch-off, resulting in larger droplets formation.  The increase in inertial force of the dispersed phase with the constant shear force exerted by the continuous phase delays the droplet pinch-off, leading to the accumulation of more dispersed volume, resulting in the increasing the droplet length as shown in Figure \ref{f11}a similar trend were found in experimental work by \citet{liu2022formation}. In addition to increasing the length of the droplet, the droplet expands in all directions, the inertial and shear force of the continuous phase less significant, resulting in a decrease in the droplet velocity while increasing the Weber number shown in Figure \ref{f11}b. The liquid film thickness occupied by the microchannel is continuously decreases with increasing the dispersed flow rate, as shown in Figure \ref{f11}c, and there is an increase in pressure difference between the dispersed phase inlet and microchannel outlet for an increase in dispersed phase flow rate, shown in figure \ref{f11}d. This increases hydrodynamic pressure resists the shear force from the continuous phase, leading to dispersed phase growth followed by delayed pinch-off results in larger droplets. However, the pressure difference is less significant than with the effect of the continuous phase flow rate.

The velocity distribution in the radial direction for higher CMC concentration with increasing dispersed phase flow rate is shown in Figure \ref{f12}a. From the wall to the microchannel center, radial velocity increases with increasing Weber number. However, the velocity profile is the same from the channel wall to 500 $\mu m$ irrespective of the dispersed phase flow rate because the dispersed phase only dominates in the center of the microchannel, whereas the continuous phase is constant for all the cases. The effect of dispersed phase flow rate on the non-homogeneous viscosity of the CMC concentration solutions is negligible, as shown in Figure \ref{f12}b. Figure \ref{f12}c and d shows the velocity distribution inside the microchannel. From the illustration, it is observed that as the CMC concentration increases, the velocity increases. However, there is a decrease in velocity as the Weber number is increases.

\subsection{Effect of interfacial tension}
Interfacial tension force is an important parameter that governs droplet formation and its dynamics. When two immiscible fluids get in contact with each other, a small interface between the two fluids tends to be established.
In this section, the effect of interfacial tension force on droplet formation, length, velocity, liquid film thickness, and pressure profile is comprehensively investigated. The interfacial tension force is altered from 0.001 - 0.004 N/m, keeping the continuous and dispersed phase flow rate constant and the interfacial tension force is represented in terms of modified Capillary number (The ratio of viscous force to the interfacial tension force,  $C_{a}^{\prime}= \frac{K v^n D^{(1-n)}}{\sigma} $).\\
\begin{figure}[!ht] 
	\centering
	\includegraphics[width=\columnwidth]{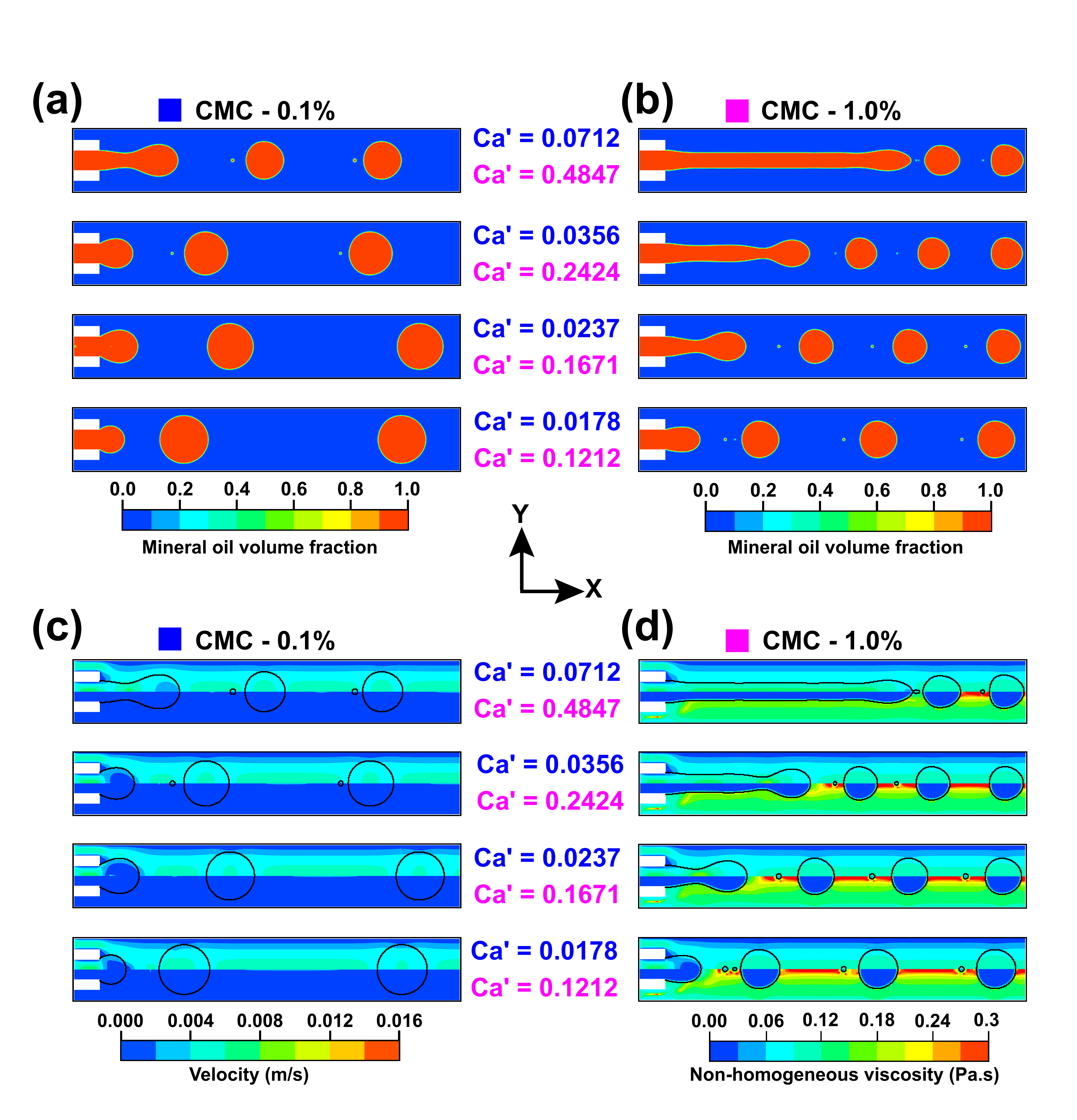} 
	\caption{ Effect of interfacial tension force on (a) mineral oil volume fraction in a CMC-0.1\%, (b) mineral volume fraction in a CMC - 1\%, (c) velocity and viscosity contours in CMC-0.1\%, and (d) velocity and viscosity contours in CMC-1\% at $Q_{c}$ = 4mL/hr and $Q_{d}$ = 3mL/hr.}
	\label{f13}
\end{figure}

\begin{figure}[!ht] 
	\centering
	\includegraphics[width=\columnwidth]{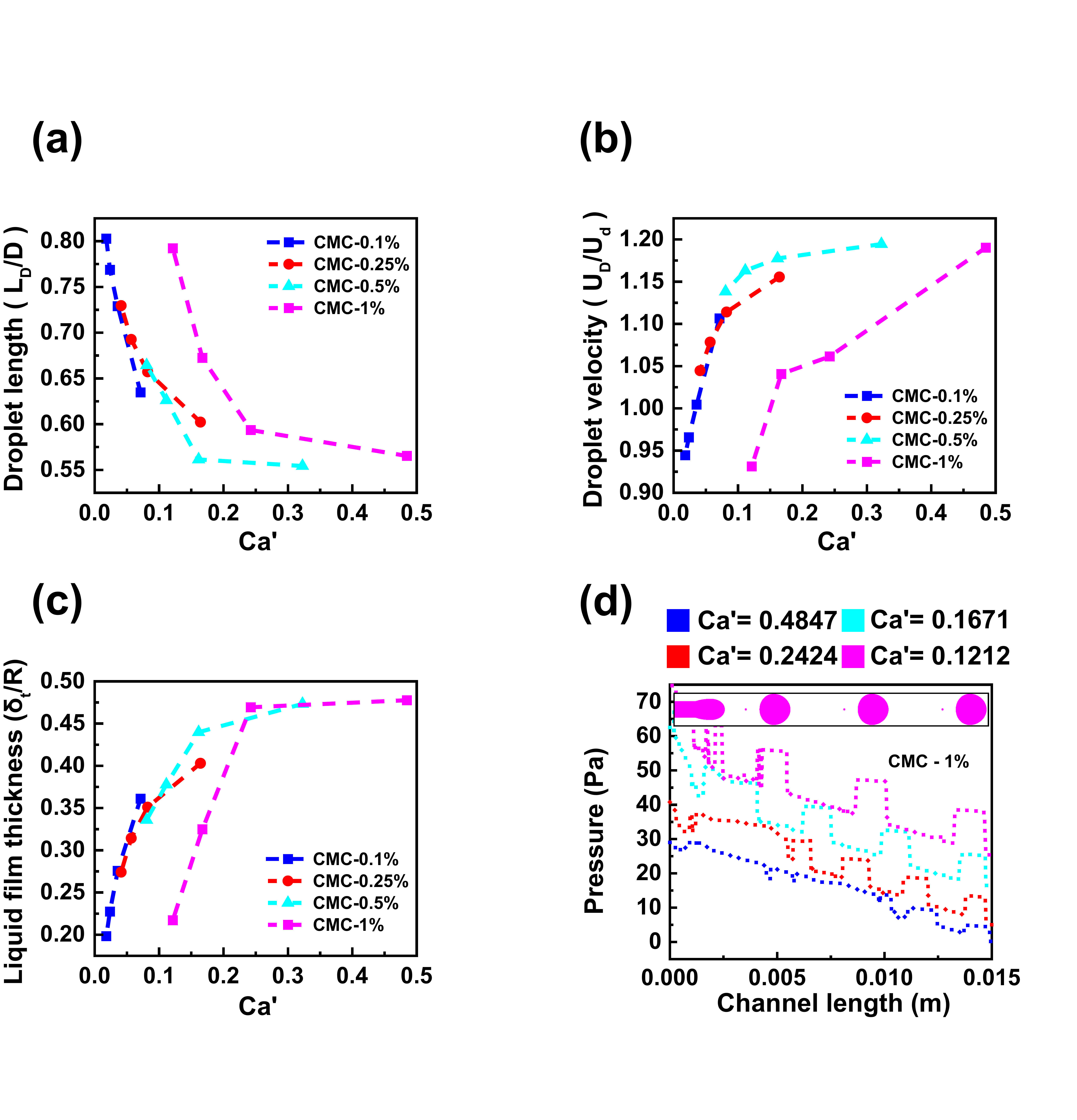} 
	\caption{ Effect of interfacial tension force on (a) non-dimensional droplet length, (b) non-dimensional droplet velocity, (c) non-dimensional liquid film thickness, and (d) pressure profile along the microchannel length for CMC- 1\% at fixed $Q_{c}$ = 4mL/hr and $Q_{d}$ = 3mL/hr.}
	\label{f14}
\end{figure}

\begin{figure}[!ht] 
	\centering
	\includegraphics[width=\columnwidth]{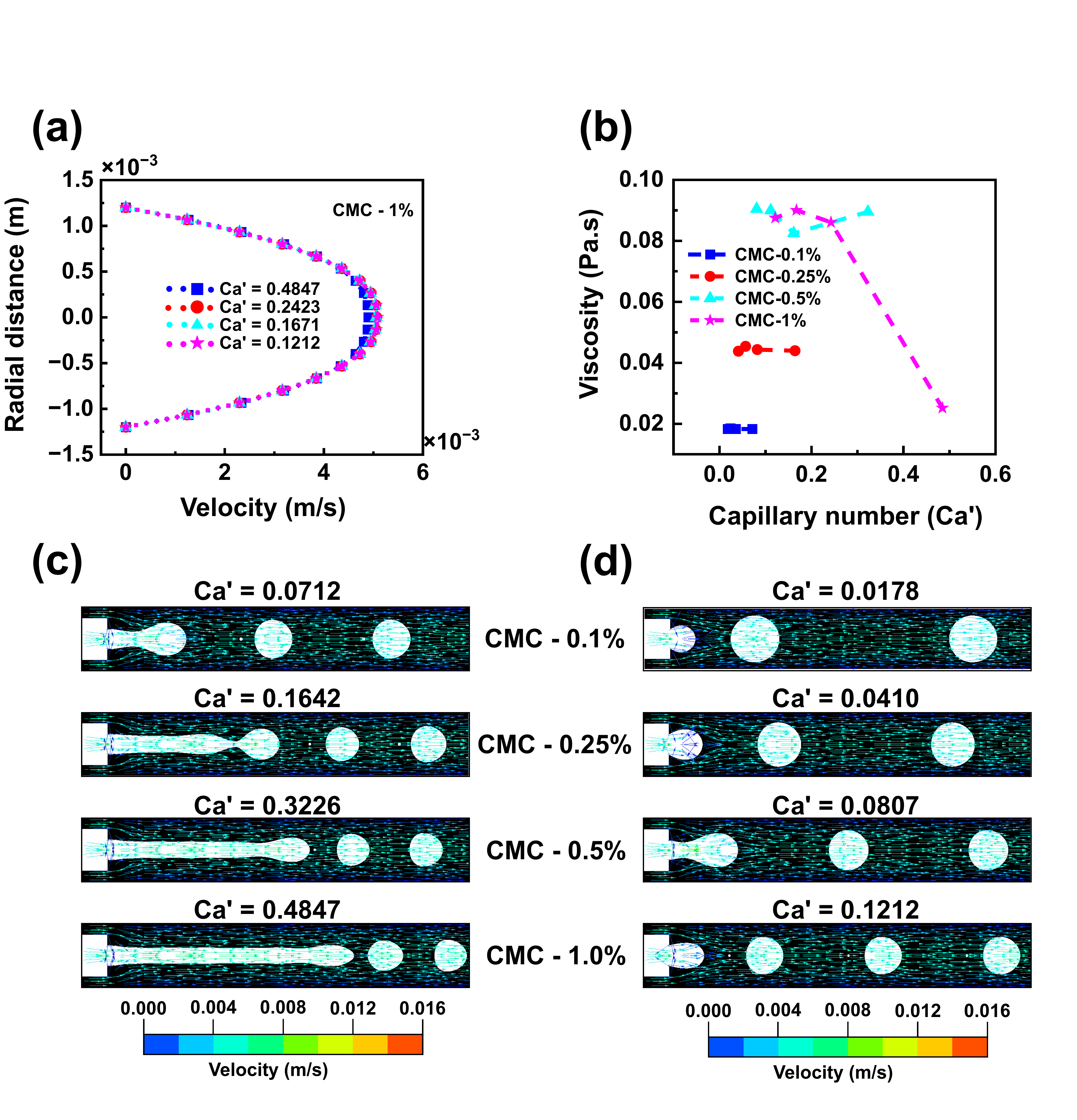} 
	\caption{ Effect of interfacial tension force on (a) radial velocity, (b) non-homogeneous viscosity, (c) velocity vector contours at lower interfacial tension force, and (d) velocity vector at higher interfacial tension force at fixed $Q_{c}$ = 4mL/hr and $Q_{d}$ = 3mL/hr.}
	\label{f15}
\end{figure}
It is observed in Figure \ref{f13}a, the droplet formation position for a CMC-0.1\% solution is shifted backward as the interfacial tension force is increases due to the lower shear force. Whereas, for CMC-1.0\%, the droplet formation is near the microchannel outlet due to higher shear force at lower interfacial tension force as shown in Figure \ref{f13}b. However, as the interfacial tension increases due to higher cohesive force acting against the shear force, the droplet position gradually shifts backward in all the CMC concentration solutions. In addition to the droplet formation location, the droplet flow regimes are examined. It is observed that for CMC-0.1\% solution concentration, the droplets are in a dripping regime, and for CMC-1\% solution, which has higher shear force at lower interfacial tension, resulting in the jetting regime to dripping with increasing the interfacial tension. The increase in  interfacial tension resulting, the force resisting the deformation is higher and resists the dispersed phase from the stretching. As a result the pinch-off position shifting from the downstream to upstream with larger droplets.  \\

Figure \ref{f13}c and d shows the velocity and viscosity contour for CMC-0.1\%  and CMC-1.0\% solutions. As the concentration increases, there is a change in viscosity in the bottom portion of the contours due to shear force. However, in both contours, the velocity contours have insignificant change due to the inertial force from the continuous and dispersed phases being constant. In Figure \ref{f14}a, non-dimensional droplet length ($L_{D}/D$) is plotted concerning modified Capillary number ($C_{a}^{\prime}$), as the interfacial tension decreases ($C_{a}^{\prime}$ increases), the droplet length  ($L_{D}/D$) is decreases. At lower interfacial tension at the interface of the two fluids, the shear force dominates and hinders the growth of the droplet.

 As the interfacial tension increases, the effect of the shear force is minimized, resulting in the droplet's growth. 
 A similar trend in droplet length is observed in a numerical simulation by \citet{sontti2019cfd}. As the $C_{a}^{\prime}$ increases, due to weaker interfacial tension force, leading to a reduction in droplet size. This reduction in droplet size creates a larger gap between the droplet and the channel wall, resulting in an increase in the liquid film thickness. A thicker liquid film allows the continuous phase to exert higher inertial force on the droplet, which reduces the resistance experienced by the droplet as it moves downstream. Consequently, the droplet is able to accelerate, resulting in a higher velocity. The non-dimensional droplet velocity ($U_{D}/U_{d}$) and liquid film thickness ($\delta_{t}/R$) are shown in Figure \ref{f14}b, and \ref{f14}c. \\
 
  As the interfacial tension increases, the pressure drop along the microchannel is increases as shown in the Figure \ref{f14}d, due to the Laplace pressure\citep{wang2021flow} $\Delta P=\sigma\left(\frac{1}{R_1}+\frac{1}{R_2}\right)$, which is directly proportional to the interfacial tension. The radial velocity profile for CMC-1\% solution is plotted in Figure \ref{f15}a, and the radial velocity has a similar trend discussed in the previous section. The non-homogeneous viscosity of all the CMC concentration solutions is plotted in Figure \ref{f15}b shows the non-homogeneous viscosity of CMC-0.1\% and CMC-0.25\% is not much affected due to the shear rate of the continuous phase is constant. However, the non-homogeneous viscosity of CMC-0.5\% and CMC-1\% is affected. The shear rate at the interface due to higher interfacial tension is high, which decreases the viscosity of shear-thinning CMC-1\% concentration solution. Figure \ref{f15} c and d show the velocity distribution inside the microchannel, flow rate of the dispersed phase, and continuous phase are kept constant, resulting in minimal velocity changes with interfacial changes.

\subsection{Flow regime and scaling}

\begin{figure}[!ht] 
	\centering
	\includegraphics[width=\columnwidth]{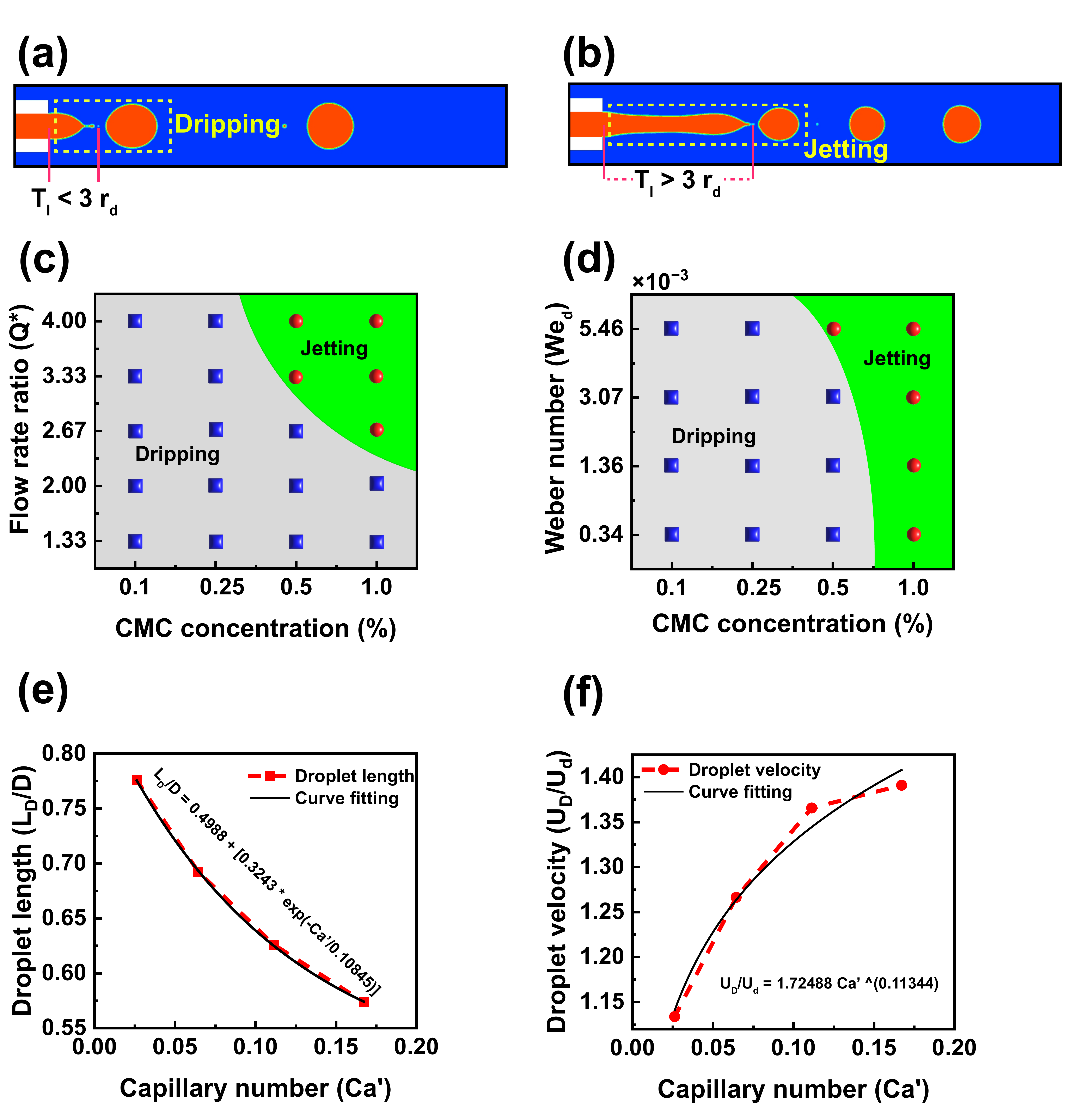} 
	\caption{Dispersed thread length (a) dripping, (b) jetting. Flow regime maps (c) effect of continuous phase flow rate at fixed $Q_{d}$ = 3 mL/hr, (d) effect of dispersed phase flow rate at fixed $Q_{c}$ = 8 mL/hr. Scaling relation (e) non-dimensional droplet length scaling relation for CMC concentration, and (f) non-dimensional droplet velocity scaling relation for CMC concentration at $Q_{c}$ = 4 mL/hr and $Q_{d}$ = 3 mL/hr.}
	\label{f16}
 \end{figure}
 In this work, we encountered dripping and jetting flow regimes. The droplets are generated at the exit of the nozzle or orifice with dispersed phase thread length less than three times the radius of the dispersed phase inlet in the dripping regime (shown in Figure \ref{f16}a), whereas, in the jetting regime, a liquid jet with an elongated thread formed which is three times greater than dispersed phase inlet radius, which eventually breaks into droplets in the middle of the channel (shown in Figure \ref{f16}b).\\
 
In this section, we have constructed two flow regime maps. In the first case, the flow rate ratio is used to represent continuous phase flow, and in the second case, the Weber number is used to represent dispersed phase flow. Two different colors are used to distinguish the flow regimes: blue represents the dripping regime, and red represents the jetting regime. For a continuous phase flow rate, the flow is in the dripping regime for CMC-0.1\% and CMC-0.25\% concentrations, whereas for CMC-0.5\% and CMC-1.0\% at higher flow rates, the flow is in the jetting regime shown in Figure \ref{f16}c. For the dispersed phase, for CMC-0.1\% and CMC-0.25\%, the flow rate is in the dripping regime, and CMC-1\% is in the jetting regime. However, for CMC-0.5\% for We = 0.00034 to 0.00307, the flow is in the dripping regime, whereas for We= 0.00546, the flow is in the jetting regime shown in Figure \ref{f16}d. Droplets generated in the dripping regime are stable droplets, whereas, in the jetting regime, the droplets are less stable comparatively.  Scaling relation for non-dimensional droplet length ($L_{D}/D$) and droplet velocity($U_{D}/U_{d}$) with the function of modified Capillary number is developed. From our CFD results, the droplet length continuously decreases with the Capillary number. As a result, the chosen exponential decay model correctly fits our data. Whereas droplet velocity continuously increases with the Capillary number, the droplet velocity trends fit closely with the power law model. The non-dimensional droplet length ($L_{D}/D$) scaling relation is developed using the exponential decay model with a coefficient of determination($R^{2} =0.99$) shown in Figure \ref{f16}e. In addition to the droplet length, the droplet velocity scaling relation is also developed using a power law model with coefficient of determination ($R^{2} =0.98$) shown in Figure \ref{f16}f.

\section{Conclusions}
A CFD model with CLSVOF interface tracking of the two-dimensional axisymmetric co-flow microfluidic device is developed to generate mineral oil droplets in a shear-thinning non-Newtonian fluid (CMC solution). The droplet break-up mechanism is discussed with effect of CMC concentration at fixed flow rates of continuous and dispersed phase flow rates. The effect of CMC solution concentration, the flow rate of the continuous and dispersed phases, and interfacial tension on the droplet are numerically examined in detail. The droplet size, velocity, liquid film thickness, and formation frequency are investigated for all the cases. It is clear from the results that the droplet length decreases with the increase in concentration and flow rate of CMC solution. However, with the increase in dispersed phase flow rate and interfacial tension, the droplet length also increases, and eventually, the liquid film thickness is opposite to the droplet length in all the cases. 

At the lower CMC concentration, the flow is in the dripping regime for all the flow rates of the dispersed and continuous phases. However, at higher flow rates and concentrations of CMC solution, the flow is in the jetting regime, whereas at higher CMC concentration for all the flow rates of dispersed phase flow rate, the flow is in the jetting regime. Increasing interfacial tension due to resistance provided by the interface increases droplet formation time and results in a similar increase in dispersed phase flow rate. The scaling relation for droplet length and droplet velocity with respect to the concentration of shear-thinning fluid is developed with the function of modified Capillary number. The fundamental aspects of the non-newtonian fluid flow and its significant role in droplet formation in a circular co-flow microfluidic device provide insightful guidance to lab-on-a-chip and industrial applications.

In addition to these findings, the results have significant implications for real-world applications such as drug delivery, material synthesis, and other microfluidic technologies where precise droplet size control is crucial. The ability to manipulate droplet formation dynamics in shear-thinning fluids can be applied to optimize processes in these fields. Future research could focus on investigating other shear-thinning fluids with different rheological properties, and validating the numerical results with experimental data. Further exploration of droplet control in more complex geometries or under varying flow conditions would enhance the practical applicability of this work.



\subsection*{Author Contributions}
M.Jammula: conceptualization (lead); methodology (lead); planned and performed the simulations (lead); model validation (lead); formal analysis (lead); software (lead); visualizations (lead);writing-original draft (lead); writing-review and editing(lead). S.G.Sontti: conceptualization (lead); methodology(supporting); data interpretation (supporting); project administration (lead); writing-review and editing (lead); resources(lead); and supervision (lead).

\section*{Declaration of competing interest}
\noindent The authors declare that they have no known competing financial interests or personal relationships that could have appeared to influence the work reported in this paper.

\section*{Acknowledgement}
\noindent 
This work is supported by the IIT Dharwad through the Seed Grant and Networking Fund (SGNF) scheme.

\section*{Data availability}
\noindent The data supporting this study's findings are available from the corresponding author upon reasonable request.

\cleardoublepage
\section*{Nomenclature}
\begin{longtable}{l p{12cm}}
$B_{o}$ & Bond number ($\Delta \rho g D^2 / \sigma$)\\
$W_{e}$ & Weber number ($\frac{\rho v^2 d_{d}}{\sigma}$)\\
$C_{a}^{\prime}$ & modified Capillary number ($\frac{K v^n D^{(1-n)}}{\sigma} $)\\
$g$ & gravitational force (m/$s^{2}$)\\
$L$ & length of the microchannel (m)\\
$D$ &  channel diameter (m) \\
$R$ & channel radius (m)\\
$r_c$ & continuous phase inlet radius (m)\\
$r_d$ & dispersed phase inlet radius (m)\\
$L_D$ & length of the droplet (m)\\
$H_D$ & height of the droplet (m)\\
$U_{D}$ & droplet velocity (m/s)\\
$U_{d}$ & dispersed phase velocity (m/s)\\
$K$ &  consistency index (Pa.$s^n$)\\
$n$ & power law index\\
$t$ & time (sec)\\
$Q_c$ & continuous phase flow rate ($m^{3}/s$)\\
$Q_d$ & dispersed phase flow rate ($m^{3}/s$)\\
$Q^{*}$ & flow rate ratio ($Q_c/Q_d$)\\
$\vec{v}$ & velocity (m/s)\\
$p$ & pressure (Pa)\\
$\vec{F}$ & surface tension force (N/m)\\
$\vec{\zeta}$ & position vector\\
$y$ & shortest distance \\
$H(\Psi)$ & Heaviside function\\
$\omega$ & interface thickness (m)\\
$\delta(\Psi)$ & Dirac delta function\\
$\mu_{eff}$ & effective viscosity\\
$D.I$ & droplet deformation index\\

\end{longtable}
\section*{Greek symbols}
\begin{longtable}{l p{12cm}}

$\mu$ & non-homogeneous viscosity (Pa.s)\\
$\dot{\gamma}$ & shear rate (1/s)\\
$\delta$ & liquid film thickness (m)\\
$\rho$ & density (kg/$m^{3}$)\\
$\sigma$ & interfacial tension force (N/m)\\
$\Psi$ & level set function\\
$\kappa$ & radius of curvature\\
$\theta$ & contact angle ($^{\circ}$)\\

\end{longtable}
\section*{Subscripts}
\begin{longtable}{l p{12cm}}

$D$ & droplet\\
$c$ & continuous phase\\
$d$ & dispersed phase\\

\end{longtable}

\section*{Supporting Information}

Comparison of LS function discretization schemes: first-order upwind and second-order upwind methods; various CMC concentrations; corresponding non-dimensional droplet length.

\cleardoublepage
\bibliography{achemso-demo}

\end{document}